\documentclass[article,usenatbib]{mnras}

\usepackage{graphicx}
\usepackage{bm}
\usepackage{amsfonts}
\usepackage{latexsym}
\usepackage[latin1]{inputenc}
\usepackage[tbtags]{amsmath}
\usepackage{epsfig}
\usepackage{amsbsy}
\usepackage{psfrag}
\usepackage{mnfixes}
\usepackage{cleveref}

\usepackage{color}

\usepackage{cancel}

\def\x{{\mathrm{x}}}
\def\y{{\mathrm{y}}}

\def\sf{{\mathrm{F}}}
\def\n{{\mathrm{n}}}
\def\b{{\mathrm{b}}}

\def\S{{\mathrm{S}}}

\def\n{{\rm n}}
\def\p{{\rm p}}
\def\e{{\rm e}}

\def\n{{\rm n}}
\def\p{{\rm p}}
\def\e{{\rm e}}

\def\e{{\rm e}}

\def\S{{\cal S}}

\def\be{\begin{equation}}
\def\ee{\end{equation}}
\def\beq{\begin{equation}}
\def\eeq{\end{equation}}
\def\bea{\begin{eqnarray}}
\def\eea{\end{eqnarray}}

\def\bear{\begin{eqnarray}}
\def\eear{\end{eqnarray}}

\begin{document}

\title[Physics of magnetohydrodynamics]{The physics of non-ideal general relativistic magnetohydrodynamics}

\author[N. Andersson et al] {N. Andersson\thanks{E-mail:na@maths.soton.ac.uk}$^1$,  I. Hawke$^1$, T. Celora$^1$ and G.L. Comer$^2$ \\ \\
$^1$ School of Mathematics and STAG Research Centre, University of Southampton, Southampton SO17 1BJ, UK \\
$^2$  Department of Physics, Saint Louis University,
St Louis, MO 63156-0907, USA}

\maketitle

\date{\today}

\begin{abstract}
We consider a framework for  non-ideal magnetohydrodynamics in general relativity, paying particular attention to the physics involved. The discussion highlights the connection between the microphysics (associated with a given equation of state) and the global dynamics (from the point of view of numerical simulations), and includes a careful consideration of  the  assumptions that lead to ideal and resistive magnetohydrodynamics. We pay particular attention to the issue of local charge neutrality, which tends to be assumed but appears to be more involved than is generally appreciated. While we do not resolve all the involved issues, we highlight how some of the assumptions and simplifications may be tested by simulations. The final formulation is consistent, both logically and physically, preparing the ground for a new generation of models of relevant astrophysical scenarios.
\end{abstract}


\section{Context}

Electromagnetic phenomena are central to neutron star astrophysics, with issues ranging from the formation and evolution of the star's internal magnetic field through to the elusive pulsar emission mechanism and the violent dynamics associated with supernova core collapse and binary mergers. The intimate connection between highly dynamical events and  powerful observed gamma-ray bursts provides ample motivation to improve the available simulation technology. In this respect, there has been notable progress towards realistic numerical simulations of neutron star mergers in full nonlinear general relativity (with the live spacetime required by Einstein's theory, see \citet{2017RPPh...80i6901B,2020GReGr..52..108B}  for reviews). In particular, given the problem we focus our attention on here, there has been interesting recent work on the role of the magnetic field. Most current efforts remain within the regime of ideal magnetohydrodynamics (see for example \citet{2021CQGra..38h5021C}), but there have also been  attempts to account for non-ideal effects, like resistivity and viscous dissipation \citep{2009MNRAS.394.1727P,2013PhRvD..88d4020D,2020MNRAS.491.5510W}. Steps in this direction are important as they take us towards a more detailed implementation of the physics, which is always desirable. In particular, we need to be able to quantify to what extent these (complicating) aspects may leave an observational signature. If they do not, then we can ``get away'' with a simpler treatment.  The argument is straightforward but it raises a number of thorny issues. Not only do we need a better handle on what the input physics should be, we also need to understand to what extent simulations are able to faithfully represent these aspects. Neither of these issues are trivial.

An important part of the discussion links  the microphysics (represented by the matter equation of state) to the large-scale dynamics. This inevitably involves considering the composition and state of matter, as well as different possible ``flows'' (associated with heat, charge currents and possibly superfluidity) that enter the problem. The recent work of \citet{2017CQGra..34l5001A,2017CQGra..34l5002A,2017CQGra..34l5003A} (see also \citet{2021FrASS...8...51A}) represents a coherent effort in this direction, outlining a flexible multi-component framework (at the level of fluid dynamics) that allows us to represent different aspects of the problem. The  results set the scene for more detailed considerations by providing both a fibration perspective---suitable for the local fluid dynamics  \citep{2017CQGra..34l5002A}---and a foliation description, geared towards spacetime simulations \citep{2017CQGra..34l5003A}. 
This paper aims to clarify the connection between the two pictures. Adding context and depth to the previous work---paying particular attention to issues relating to the assumptions associated with magnetohydrodynamics and local charge neutrality on different relevant scales---our discussion takes us another couple of steps towards realism. 

Throughout the discussion, we adopt the convention that spacetime  indices are represented by $a,b,c,...$ while $i,j,k, ...$ are spatial indices in a chosen coordinate frame. The Einstein summation convention is assumed for both sets.

\section{The equations of electromagnetism}

It is natural to begin by considering the equations that govern the electromagnetic degrees of freedom. The 3+1 form for Maxwell's equations is well known \citep{baum}, but this exercise is nevertheless useful as it establishes the procedure we use for the fluid dynamics. It also offers an opportunity to highlight how the issue of charge neutrality  leads to (potentially unavoidable) uncertainties in the modelling.

\begin{figure}
\begin{center}
\includegraphics[width=\columnwidth]{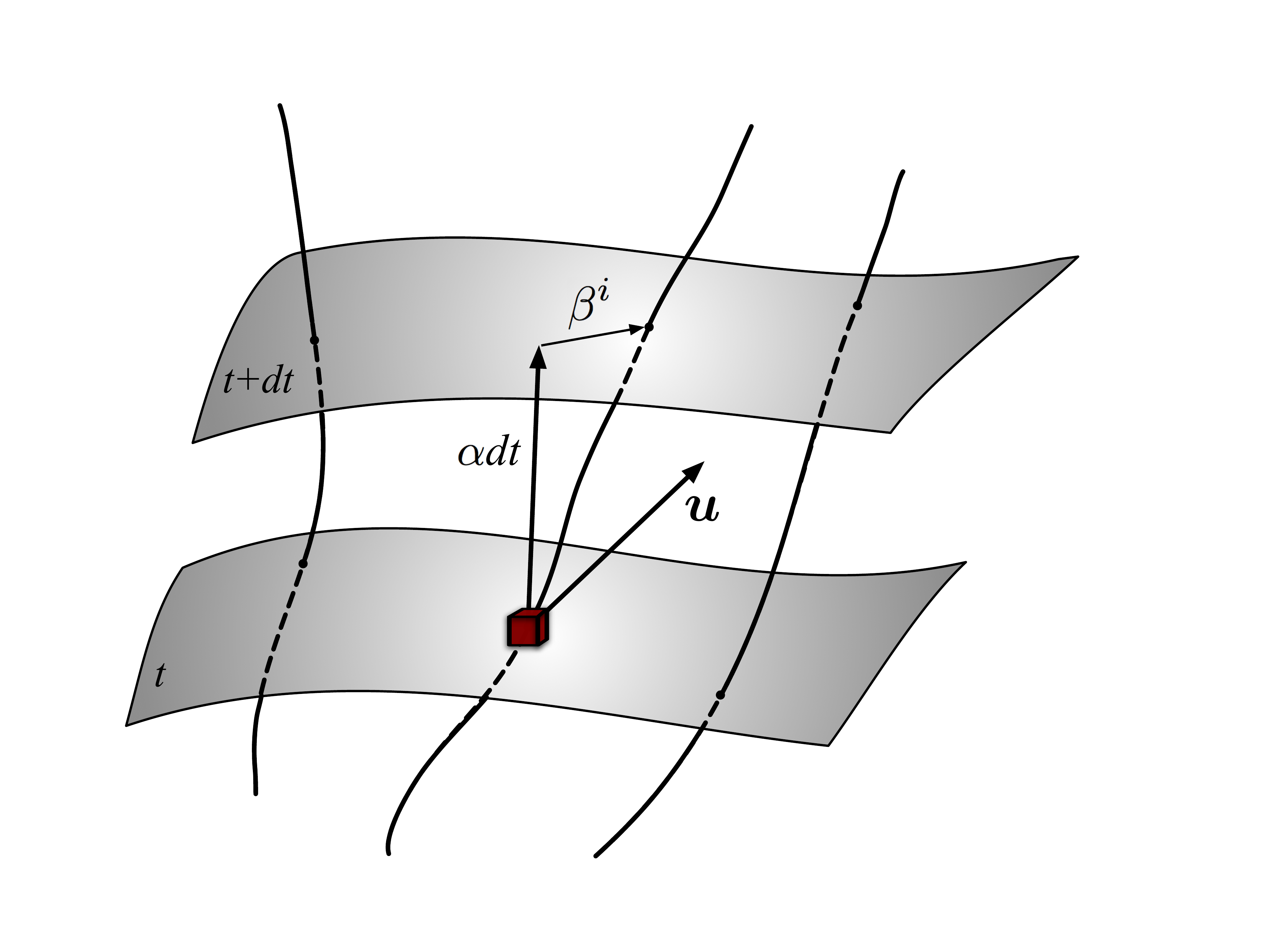}
\end{center}
\caption{Illustrating the connection between two formulations of the relativistic fluid problem. The fibration approach, which focuses on the worldline associated with a given fluid element (and a  four velocity $\boldsymbol u$ with components $u^a$), provides a natural description of the microphysics and issues relating to the local thermodynamics. Meanwhile, a spacetime foliation, based on the use of spatial slices and normal observers (depending on a lapse $\alpha$ and a shift vector $\beta^i$), is typically used in  numerical simulations. In order to ensure that the local physics is appropriately implemented in simulations, we need to understand the translation between the two descriptions.} 
\label{fibrate}
\end{figure}

Key to the discussion is the relation between the local physics---encoded by a matter equation of state---and the variables used in a numerical  simulation (see \cref{fibrate}). In particular, we need to understand how the evolved variables connect with the microphysics and the local thermodynamics. In a fully nonlinear/nonequilibrium system this is a  challenging problem and we do not expect to resolve all the involved issues here (especially those linked to small scale turbulence, see \citet{2021arXiv210701083C} for a recent discussion and references to the relevant literature). Having said that, we will demonstrate how we can make progress by making physically reasonable approximations.

\subsection{Maxwell's equations}

To set the scene,  recall that the 3+1 decomposition used in nonlinear gravity simulations \citep{baum,2013rehy.book.....R,LivRev} involves a set of spatial hypersurfaces and (Eulerian) observers associated with the corresponding normal, $N^a$. The spacetime metric is given by 
\be
ds^2 = - (\alpha^2 - \beta^2) dt^2 + \gamma_{ij} (\beta^i dx^j + \beta^j dx^i) dt + \gamma_{ij} dx^i dx^j \ , 
\label{gij}
\ee
 where $\alpha$ and $\beta^i$ represent the lapse and the shift, respectively (see \cref{fibrate}). The spatial metric $\gamma_{ab}$ acts as a projection orthogonal to $N^a$ and is used to introduce a suitable  derivative in each spatial hypersurface
 \be
 D_i = \gamma_i^b \nabla_b  \ ,
 \ee
 where all free indices should be projected.
We also introduce the Christoffel symbols $ \Gamma^j_{ki}$ associated with $\gamma_{ij}$, ensuring the compatibility $D_i \gamma_{jk} = 0$.
Moreover, if   $\gamma$ is the  determinant of the induced metric, then 
\be
D_i \gamma^{1/2} = \partial_i \gamma^{1/2} - \Gamma^j_{ji} \gamma^{1/2} = 0  \ .
\label{Dgamma}
\ee

We also need the extrinsic curvature
\be
2\alpha K_{ij} = - ( \partial_t - \mathcal L_\beta)  \gamma_{ij} \ ,
\label{Kijdef}
\ee
where $\mathcal L_\beta$ is the Lie derivative along $\beta^i$. 
We then have
\be
\mathcal L_\beta  \gamma_{ij} = \gamma_{kj}D_i \beta^k + \gamma_{ik} D_j \beta^k = D_i \beta_j + D_j \beta_i \ ,
\ee
and it follows that the trace of the extrinsic curvature satisfies
\be
\alpha K = - \partial_t \ln \gamma^{1/2} + D_i \beta^i  \ .
\label{alpK}
\ee

Turning to the equations of electromagnetism, 
let us (for clarity) assume that we opt to work with the electric and magnetic fields. (An alternative description based on working with the vector potential can be found in \citet{baum}, with recent implementations discussed by \citet{2020ascl.soft04003E,2021CQGra..38h5021C}.) These are (obviously) observer dependent quantities. In terms of the Faraday tensor $F_{ab}$ an Eulerian observer (associated with $N^a$) will measure the electric field 
\be
E_a = -  N^b F_{ba} \ ,
\ee
and the magnetic field
\be
B_a =  {1 \over 2} \epsilon_{abc}F^{bc} \ .
\ee
where we have defined
\be
\epsilon_{abc} = \epsilon_{dabc}N^d \ , 
\ee
(associated with a right-handed coordinate system moving along with $N^a$). 
This then leads to
\be
F_{ab} =   N_{a} E_{b} - N_b E_a  + \epsilon_{abc} B^c \ .
\label{Faraday_appendix}\ee
Clearly, the electric and magnetic fields are orthogonal to $N^a$ (by construction)  
and so each will have only three components (as expected).

The equations that govern the electromagnetic field are (obviously) well known, so we simply draw on the results from \citet{2017CQGra..34l5003A}.
First of all, we need to introduce the Eulerian frame decomposition of the charge current 
\be 
j^a = \hat \sigma N^a + \hat J^a \ , \qquad \hat J^a N_a = 0 \ .
\ee
The charge current actively generates and sustains the electromagnetic field. Yet,  much of the astrophysics literature assumes that the charge current plays a more passive role. This step---one of the assumptions that leads to magnetohydrodynamics---is important. It effectively reduces the problem from a multi-component plasma (see  \citet{2012PhRvD..86d3002A} for discussion and pointers to the relevant literature) to a ``single-fluid'' model that is easier to work with. The arguments in favour of this strategy are well developed in  non-relativistic physics \citep{1999stma.book.....M}, but the extension to general relativity tends to be made by analogy rather than in-depth analysis. Given this, we will pay particular attention to issues related to the charge current in the following.  

 Maxwell's equations follow from, first of all
 \be 
 \nabla_b F^{ab} = \mu_0 j^a\ ,
 \ee
which, from the foliation perspective, leads to a relation between the divergence of the electric field and the charge density $\hat \sigma$ 
\be
D_i E^i = \mu_0  \hat \sigma \ ,
\label{divE}
\ee
where $\mu_0$ is the magnetic permeability. 
Here, and in the following, we use the same matter quantities as in \citet{2017CQGra..34l5003A} and---in order to make clear the distinction---retain the convention of using hats to denote matter quantities measured by Eulerian observers (while the corresponding electric and magnetic fields are given as capital letters). 
It is also worth noting that, since  we are assuming $c^2=1/\mu_0 \varepsilon_0 = 1$ we recover the standard form for Gauss' law: 
\be
D_i E^i ={  \hat \sigma \over \varepsilon_0} \ .
\ee
We also have an evolution equation for the electric field\footnote{Here, and in the following, we write the evolution equations in a way that focusses on the physics involved, rather than the flux-conservative form required for a numerical implementation. The translation between the descriptions is, however, standard and should not present any particular difficulties.}
\be
\left( \partial_t - \mathcal L_\beta\right) E^i - \epsilon^{ijk} D_j (\alpha B_k) + \alpha \mu_0 \hat J^i = \alpha K E^i \ .
\label{dtE}
\ee

The second pair of Maxwell equations follow from
\be 
\nabla_{[a}F_{bc]}=0 \ .
\ee
Given the absence of magnetic monopoles, we have
\be
D_i B^i = 0  \ ,
\label{divB}\ee
while the magnetic field evolves according to 
\be
\left( \partial_t - \mathcal L_\beta\right) B^i + \epsilon^{ijk} D_j (\alpha E_k) = \alpha K B^i \ .
\label{dtB}
\ee

It is important to keep in mind that, in practice, these equations  refer to a computational cell on a specified numerical grid. There is no actual observer that moves through spacetime with four-velocity $N^a$. This is obvious, but deserves emphasis as a typical grid resolution involves a fluid ``box'' that is much larger than the fluid ``elements'' of the underlying fluid model. The relevance of this should (hopefully) become clear as we proceed.

 \subsection{Towards magnetohydrodynamics}
 
Astrophysical problems tend to be considered in the context of magnetohydrodynamics. 
As we want to understand  the  physics of the problem, it is useful to spell out how this simplifies the equations.
The argument is fairly straightforward. Let us assume that the dynamics is associated with  characteristic length- and timescales, $L$ and $T$,  leading to  an associated velocity $V\sim L/T$ (noting that $K\sim 1/T$  and ${\mathcal L_\beta} \sim 1/T$). It then follows from \cref{dtB} that 
\begin{multline}
    \underbrace{\left( \partial_t - \mathcal L_\beta - \alpha K\right) B^i}_{\sim B/T} + \underbrace{\epsilon^{ijk} D_j (\alpha E_k)}_{\sim E/L} = 0 \\  \Longrightarrow\ B \sim  {E/V}\ .
\end{multline}
Similarly, it follows from \cref{dtE} that 
\be
\underbrace{\left( \partial_t - \mathcal L_\beta - \alpha K\right) E^i}_{\sim E/T} - \underbrace{\epsilon^{ijk} D_j (\alpha B_k)}_{\sim B/L} + \alpha \mu_0 \hat J^i = 0 \ .
\ee
We see that that we can safely neglect the displacement current (the first term) as long as $V^2 \ll 1\ (=c^2)$.  In effect, we have a low frequency/velocity approximation. Leaving out the corresponding contribution to \cref{dtE} we arrive at the familiar relation from magnetohydrodynamics:
\be
 \epsilon^{ijk} D_j (\alpha  B_k )\approx \alpha  \mu_0 \hat J^i \ .
 \label{mhdF}
 \ee
As the charge current is slaved to the magnetic field, we have effectively removed a dynamical degree of freedom from the problem. 

The argument is (of course) standard in flat space (and the usual Cartesian coordinates), where we have $\alpha = 1$, $\beta^i=0$ and $\gamma=1$ in the various equations and the covariant derivatives reduce to partials \citep{bellan,1999stma.book.....M}. The suggested extension to the curved spacetime setting is intuitive, but we may have to tread a bit more carefully. Clearly, the scaling argument relies on the assumption that the gauge (the choice of lapse and  shift) does not impact on the scaling, but there is no guarantee that it could not. For example,  if we consider $\beta^i$ as (effectively) a velocity then we clearly have to restrict it to be small (in a suitable sense). However, there is no reason why one would not be ``allowed'' to consider gauges with a large enough $\beta^i$ that the argument is messed up. With a non-trivial choice of the shift vector, one would at least have to consider the possibility that this impacts on the dynamics. In a similar fashion, the lapse $\alpha$ may affect the assumed scalings. In essence, we have to apply the  magnetohydrodynamics approximation with some level of caution, as the logic leading to \cref{mhdF} inevitably involves a degree of  \emph{gauge dependence}. One might consider testing the result by actual simulations, which ought to be fairly straightforward. Having said that, perhaps the most natural attitude is to pragmatically assume that any choice of gauge that breaks the logic is likely to be somewhat artificial, ignore the issue and move on. 
 
 Let us see where this takes us. We now have a different problem. By effectively working with the pre-Maxwell  form of \cref{dtE} (leaving out the displacement current) we cannot solve \cref{dtB} without providing $E^i$. We need an additional relation between the electric and magnetic fields. This is where the issue of the conductivity (effectively Ohm's law) enters the discussion. In a perfect conductor, where charges easily flow, one would expect the electric field to ``short out'' as the matter becomes locally charge neutral. As this argument brings in the local physics associated with a given fluid element, let us change perspective and consider the problem from that point of view.

\subsection{The local view}\label{subsec:LocalView}

The local description introduces a different observer, with four velocity $u^a$ (as in \cref{fibrate}), associated with the fluid motion. Connecting to the Eulerian observers from the foliation picture, we have 
\be
u^a = W \left(N^a + \hat v^a \right) \ , \qquad N_a \hat v^a = 0 \ , 
\label{fvel}
\ee
with $\hat v^a$ the relative velocity between the two frames and  the Lorentz factor
\be
W = \left( 1-\hat v^2\right)^{-1/2}\ , \qquad  \hat v^2 = \gamma_{ij} \hat v^i \hat v^j \ .
\ee

The electric field measured by the fluid observer follows from
\be
e_a = -u^b F_{ba} \ ,
\ee
leading to  
\be
e_a =W \left[ E_a + \epsilon_{abc} \hat v^b B^c + N_a \left( \hat v^b E_b \right) \right] \ .
\ee
It is evident that, in general, the electric field inferred by the local observer has a component parallel to $N^a$
\be
e^\parallel = - e^a N_a = - W \left(\hat v^b E_b\right) \ ,
\ee 
as well as an orthogonal piece
\be
e_a^\perp = W  \left( E_a + \epsilon_{abc} \hat v^b B^c \right) \ .
\label{mhd1}
\ee

Let us now bring in the assumption that the local electric field vanishes, as one would expect from the standard (Debye) screening argument in a perfect conductor. Specifically, if we let 
\be
e_a^\perp = 0 \quad \Longrightarrow \quad  E_a + \epsilon_{abc} \hat v^b B^c= 0 \ ,
\label{EMHD}
\ee
then is easy to see that we also have $e^\parallel=0$. If we combine \cref{EMHD} with \cref{dtB} and \cref{mhdF} we have all the relations we need to solve the problem. This is ideal magnetohydrodynamics. The arguments are, of course, standard but---as we will soon see---we need to pay attention to a number of subtle points. In particular, we need to carefully consider the issue of charge neutrality.  

\subsection{Charge neutrality}

In addition to invoking the low-frequency argument to simplify the Maxwell equations, text-book magnetohydrodynamics involves an assumption of local charge neutrality. This is, effectively, the (previous) argument that the local electric field vanishes.  The result follows immediately from the flat-space version of the Gauss law \cref{divE}. If the electric field vanishes, then so does the local charge density. 
However, the problem turns out to be more involved---and interesting!---in the relativistic setting. 

In order to see why this is the case, we may start from the  charge current, which has to satisfy
\be
\nabla_a j^a=0\ ,
\label{currcon}
\ee
in order to ensure electromagnetic gauge invariance. This is not an independent result---it can be obtained from the Maxwell equations \citep{1982MNRAS.198..339T}---but it is nevertheless useful to consider it separately. In terms of the Eulerian variables, we have
\be
\left( \partial_t - \mathcal L_\beta\right)\left( \gamma^{1/2} \hat \sigma \right)+ D_i \left( \gamma^{1/2} \alpha \hat J^i \right) = \gamma^{1/2} \hat \sigma D_i \beta^i \ .
\label{chargecon}
\ee
Invoking \cref{mhdF}, we see that 
\be
 \mu_0 D_i \left( \gamma^{1/2} \alpha \hat J^i \right) \approx \epsilon^{ijk} D_i D_j \left( \gamma^{1/2} \alpha B_k\right) = 0 \ , 
\ee
which leaves us with
\be
\left( \partial_t - \mathcal L_\beta\right)\left( \gamma^{1/2} \hat \sigma \right) = \gamma^{1/2} \hat \sigma D_i \beta^i \ .
\label{chargecon2}
\ee

In flat space (e.g. special relativity, with $\beta^i=0$ and $\gamma=1$, as before) we immediately arrive at the usual argument for quasi-neutrality:
\be
\partial_t \hat \sigma = 0 \ .
\ee
If a system starts out with $\hat \sigma = 0$ (e.g. due to screening on length scales of relevance for the evolution) then this condition is preserved as time marches on. 

In the curved spacetime case the problem is more subtle since  simulations tend to involve non-trivial choices for the lapse and  shift (and the logic obviously breaks if \cref{mhdF} does not hold, as discussed earlier), but it might still seem reasonable to argue that an evolution initiated with a uniformly vanishing charge density should remain charge neutral (in this sense). In principle, we are ``allowed'' to assume $\hat \sigma=0$ (as in the example used as illustration by \citet{2017CQGra..34l5003A}), but the question is if this is a ``sensible'' thing to do\footnote{As a slight aside, it is worth noting that we could always, in principle, construct a gauge such that \cref{chargecon} preserves any initial $\hat \sigma$. All we need to do is set
$$
\hat \sigma \beta^i = \alpha J^i \ .
$$
Of course, this only works for a non-vanishing $\hat \sigma$, as otherwise $\alpha =0$ and time would not progress. Moreover, the suggestion is unlikely to ever be relevant as one has to reserve the gauge choices to deal with more serious issues.}. It will soon become clear that it is not.

For example, if we 
combine \cref{EMHD} with \cref{divE} we see that 
\be 
\mu_0 \hat \sigma = - D_i \left( \epsilon^{ijk} \hat v_j B_k \right) \ .
\label{hsone}
\ee
This is also a well-known result---of immediate relevance for neutron star astrophysics as it leads to the Goldreich-Julian charge density for rotating magnetospheres \citep{1969ApJ...157..869G}--- enforcing the point that we should not expect $\hat \sigma = 0$ to hold \emph{everywhere}. 

Noting this argument, let us shift the emphasis to the local charge density 
 measured by the fluid observer, $\sigma$. This follows from 
\be
j^a =  \sigma u^a +  J^a \ , \qquad u^a  J_a = 0 \ , 
\ee
leading to
\be
 \sigma = - u^a j_a = W\left( \hat \sigma - \hat v^a \hat J_a \right) \ .
\ee
That is, if the matter is \emph{locally} charge neutral (in the sense that $\sigma=0$) then the quantities measured by the Eulerian observer must satisfy
\be
\hat \sigma - \hat v^a \hat J_a  = 0 \ .
\label{neut}
\ee
The result is intuitive. A change of  observer frame impacts on measured volumes and hence the charge density and the associated current.  

So far, we have essentially summarized the standard approach to (ideal) relativistic magnetohydrodynamics. We have seen how we may represent the charge current in terms of the curl of the magnetic field as long as we ignore the displacement current. We have also seen how the condition of local charge neutrality ($\sigma=0$) enters the discussion and how the dynamics (e.g. bulk rotation) may induce an effective  large-scale charge density ($\hat \sigma \neq0$), as in \cref{hsone}. 

\subsection{Averaging from the nuclear physics scale}

From the neutron star physics point of view it seems inevitable that the condition of local charge neutrality ($\sigma=0$) should hold on the scale relevant for nuclear physics (in all reasonable settings, with a possible caveat for extended regions with mixed phases). Basically, the fact that the electrons are highly mobile makes the relevant screening length vastly smaller than the size of a typical fluid element (which must, in turn, exceed (say) the electron mean free path). Given this, local charge neutrality is (almost exclusively) assumed in modern equation of state calculations\footnote{The situation is notably different for beta equilibrium, for which the governing reactions (the Urca reactions in the case of a neutron star core) are slow enough that the system may not reach equilibrium on the time scale of (say) core collapse or neutron-star merger (see \citet{ham21} for a recent discussion).}.
Basically, we need to take the condition \cref{neut} seriously.

\begin{figure}
\begin{center}
\includegraphics[width=0.8\columnwidth]{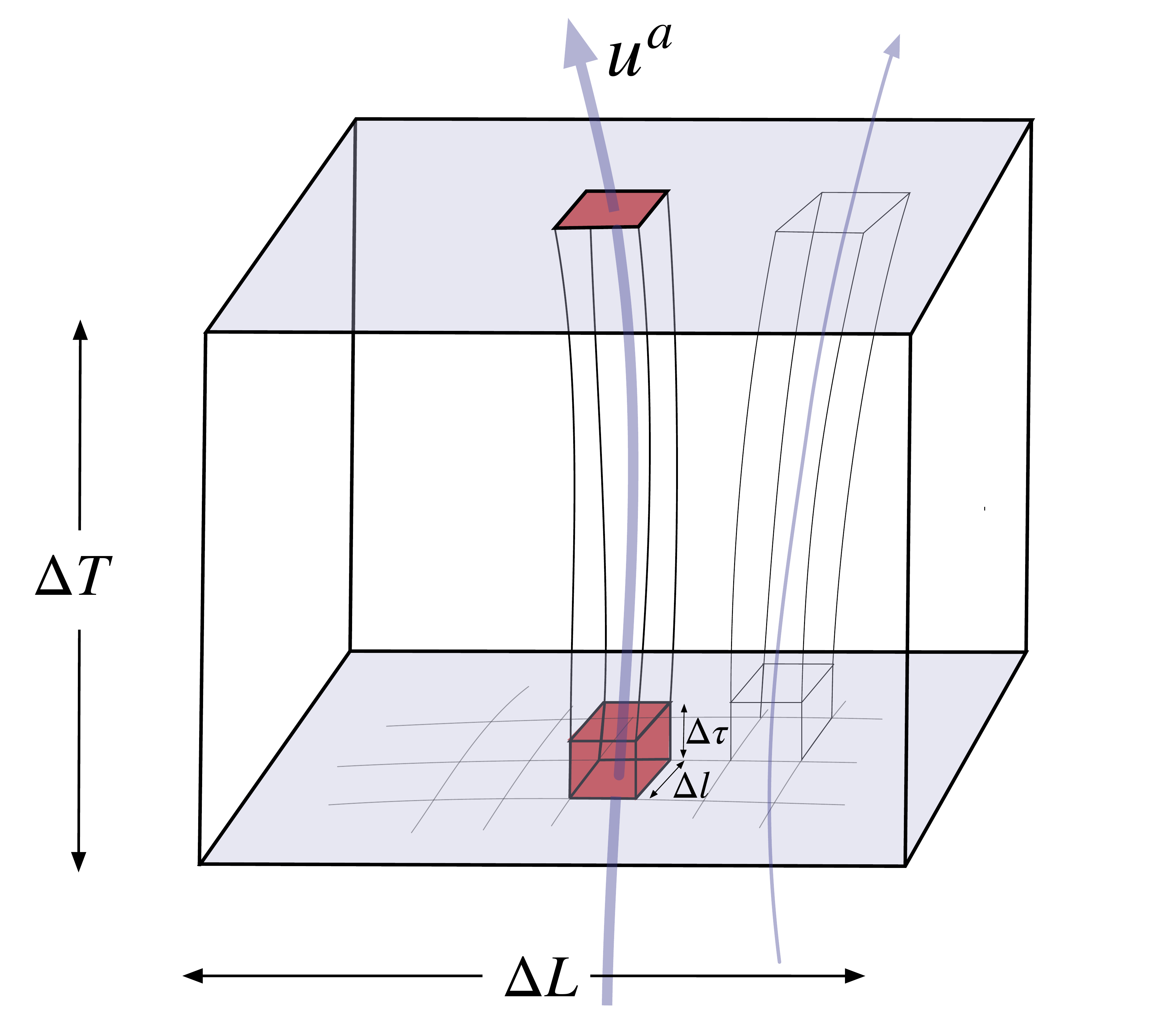}
\end{center}
\caption{A fluid element inside a much larger computational cell.} 
\label{cell}
\end{figure}

The question is on what scale we have to enforce the condition---exactly how local is local?
In order to explore this issue, let us zoom in on a fluid element from a numerical evolution, as indicated in \cref{cell} (where $\Delta L$ could be as large as 100m across even in a state of the art simulation \citep{2017RPPh...80i6901B}), to a much smaller scale (still large enough that we can meaningfully use the fluid description, leading to a typical $\Delta l$ of something like a few mm in a neutron  star core \citep{LivRev}). We want to understand the impact of the averaging from the smallest scale in the problem to the computational scale and to what extent there is a risk of key aspects being ``lost in translation''. This is a legitimate concern because, in essence, one assumes that the physics can be adequately represented by average values for the different fields. First of all, the fluid model itself is (obviously) based on the notion of averaging over a large number of particles (represented by suitable distribution functions in the more fine grained kinetic theory picture). Secondly, we average again to reach the computational scale. As we will see, both steps require careful consideration.

Let us first consider the problem at the level of individual fluid elements. Letting the local fluid frame  be represented by $u^a$, the four velocity of a fluid component exhibiting relative flow,  $u_\x^a$, is  given by \citep{2017CQGra..34l5002A}
\be
u_\x^a = \gamma_\x \left( u^a + v_\x^a \right) \ , \qquad u_a v_\x^a = 0 \ ,
\label{uxa}
\ee
where 
\be
\gamma_\x = \left( 1 - v_\x^2 \right)^{-1/2} \ .
\ee
In the problem at hand we need there to be a relative flow because the system has to sustain a charge current. If we assume that the charge carriers are electrons ($\x=\e$) and protons (p) as in a neutron star core, then \be
j^a =  \sum_{\x=\p,\e}q_\x n_\x^a = e \left(n_\p^a - n_\e^a\right) \ ,  
\ee
where $n_\x^a = n_\x u_\x^a$ and    $n_\x=-u^\x_a n_\x^a$ is the (co-moving) number density of each species, while $q_\x$ is the charge per particle (so $q_\e=-e$). It follows that, in general, we have
\be
\sigma = -u_a j^a = e (n_\p \gamma_\p - n_\e \gamma_\e) \ .
\ee

At this point we note that we need to keep track of the individual Lorentz factors, $\gamma_\x$. This may be problematic as it implies that we keep track of the individual velocities, i.e. work at the level of a multi-component plasma (which would at the very least be computationally expensive). Given this, and the fact that we want to make contact with the underlying microphysics (and the charge neutral equation of state), which is determined in dynamical equilibrium, it is natural to simplify the problem by assuming that the relative drift is sufficiently slow that we can linearise the relations to ensure that $\gamma_\x \approx 1$. At the linear drift level, all observers (e.g. comoving with either of the particle species) will agree on the number densities and the notion of charge neutrality for a given fluid element is not contentious.  %\todo{gyro radius argument?}

It is important to note that the linear drift assumption does not imply the single-fluid approximation. We still retain  the distinct flows of the system, although these are now assumed to be sufficiently close that the approximation makes sense. 

\begin{figure}
\begin{center}
\includegraphics[width=0.6\columnwidth]{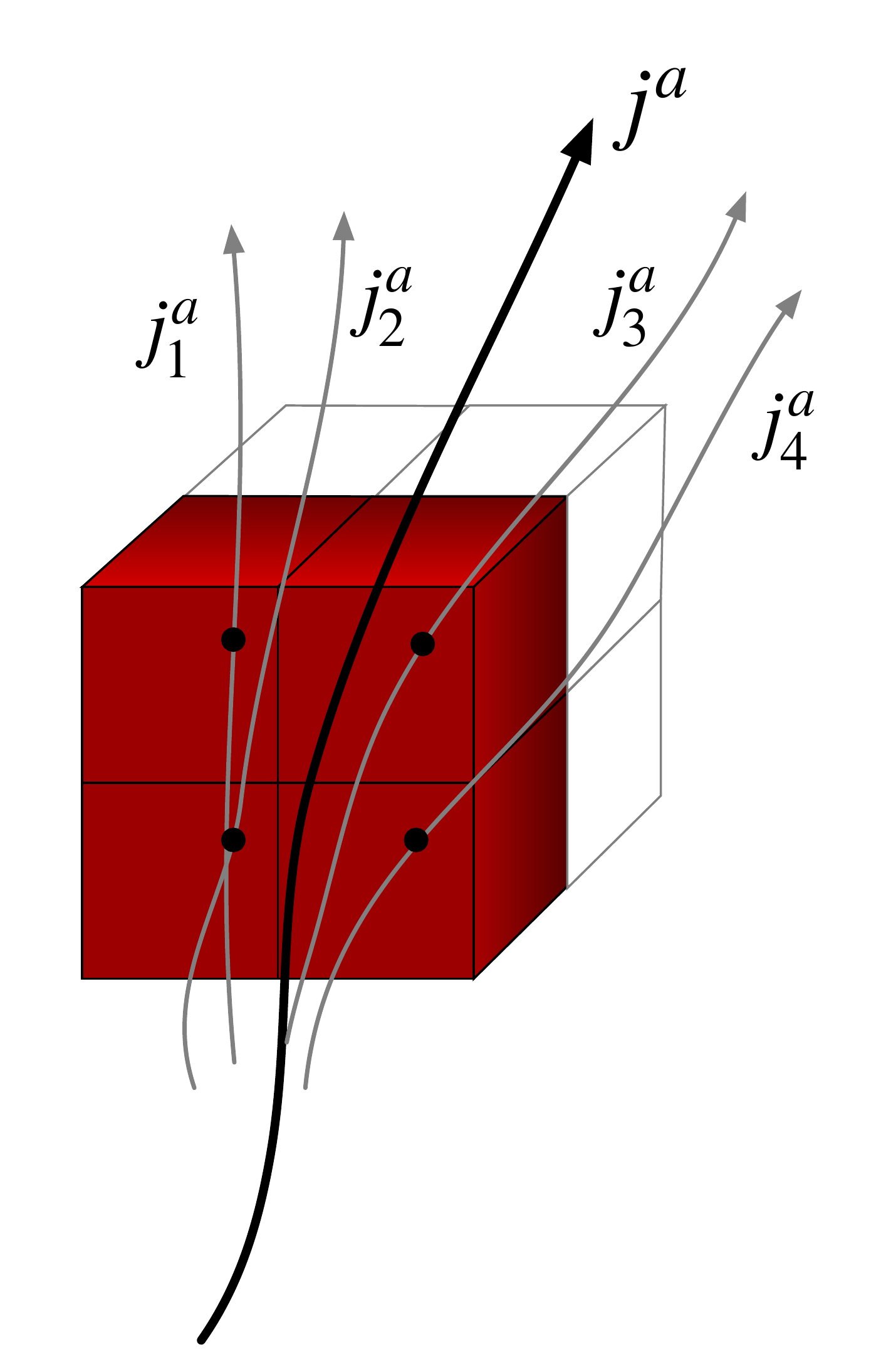}
\end{center}
\caption{A schematic illustration of the boxes within boxes used in the discussion of the averaging over charged flows.} 
\label{boxes}
\end{figure}

Let us now ask what happens if we scale the argument up to the (vastly larger) evolution scale. In effect, we consider a set of ``boxes within boxes'', illustrated in \cref{boxes}, and ask how the physics  averages as we return to the evolution scale. The main point is to illustrate that this involves unknown (perhaps even unknowable) aspects. For obvious reasons---given the context---we concentrate on the charge current. We then need 
\begin{equation}
    j^a = \sum_\x q_\x n_\x u_\x^a \approx \sum_\x q_\x n_\x (u^a+v_\x^a) = \sigma u^a + J^a\ ,
\end{equation}
where
\begin{equation}
    u^a J_a = 0 \ ,
\end{equation}
and we have (again) made use of the linear drift approximation. 
%First of all, this shows that, if we  ignore quadratic terms in $v_\x^a$ we have to ignore quadratic terms in $J^a$, as well. 
For fluid elements on the smallest scale, we impose local charge neutrality (as per the previous argument) so take $\sigma =0$. However, this condition only holds along the world line of  a particular fluid element. Suppose we consider a set of neighbouring fluid elements, as in \cref{boxes}, and ask how the argument changes as we average over a larger volume. Labelling quantities associated with each box by $N=1,2,...$ we then have the total charge current
\be
j^a = \sum_N \left[ - (u_b j^b_N) u^a + \perp^a_b j_N^b\right]  \ ,
\ee
where the projection is associated with the four velocity of the observer on the averaged scale. In general, this observer would not record a vanishing charge density (even though each $\sigma_N=0$). However, recalling the linear drift argument we have
\be
\sigma = - \sum_N u_b j^b_N \approx \sum_N (u^N_b - v^N_b) j^b_N = - \sum_N v^N_b j^b_N  \ .
\label{sigsum}
\ee 
Since each term in the sum is quadratic in small quantities it would seem consistent to ignore the contribution and take the charge current (on the larger scale) to be given by
\be
j^a =  \perp^a_b \sum_N j_N^b \ .
\ee
This argument extends to the scale on which we are carrying out the evolution, providing (some) support for the constraint from \cref{neut}.

However, one can easily come up with a counterargument. Suppose that the linear drift introduces a small scale such that $j_N^a \sim \mathcal O(\epsilon)$. Then we can easily break the logic by taking the number of boxes in \cref{sigsum} to be of $\mathcal O(1/\epsilon)$. And clearly, in scaling up from  the local fluid scale ($\sim$mm)  to the numerical evolution ($\gg$m) we are dealing with a large number of boxes. This may be problematic, but what we should replace the argument with is less obvious. The key question is if the connection between charge neutrality on the nuclear physics scale (encoded in the equation of state) breaks as we average to the evolution scale, and if so, how we account for the corresponding (subgrid) behaviour in an evolution.

\subsection{A filtering argument}

Having explained the issue, let us  consider how we might be able to  get a more quantitative handle on it. To do this we draw on the recent discussion of spacetime filtering/averaging in the context of large-eddy models of turbulence from \citet{2021arXiv210701083C}. The main idea is that, we may squeeze in a lot of physics when we scale up to the resolution of a numerical simulation. We need an effective representation of this physics, e.g. small scale fluctuations, on the resolved scale. Following the discussion of \citet{2021arXiv210701083C} we may develop the required model by describing the different physical fields according to  a ``Favre-weighted'' observer $\tilde u^a$ (constructed to ensure that the equation for baryon number conservation takes the pre-filtered form, without additional closure terms).  We then have the filtered charge current (we will not need to prescribe the actual filtering procedure, which is denoted by $\langle ...\rangle$, in order to make the argument we are interested in here)
\be
\langle j^b \rangle = \tilde \sigma \tilde u^a + \tilde J^a \ , \quad \tilde u_a \tilde J^a = 0 \ . 
\ee
Introducing the projection 
\be
\tilde \perp^a_b = \delta^a_b + \tilde u^a \tilde u_b
\ee
we have
\be
\tilde \sigma = - \tilde u_a \langle j^a \rangle
\ee
and
\be 
\tilde J^a = \tilde \perp^a_b \langle j^b\rangle \ .
\ee
The fields we would study in an evolution---on the macroscopic scale---are $\tilde u^a$, $\tilde \sigma$ and $\tilde J^a$.
Naturally, we can compare these  quantities to the filtered version of the fine-scale quantities. We then have
\be 
\langle j^a \rangle = \langle \sigma u^a \rangle + \langle J^a \rangle
\ee
and it follows that
\be 
\tilde \sigma = - \tilde u_a \left( \langle \sigma u^a \rangle + \langle J^a \rangle\right)
\ee
From this relation it is evident that, even if the small-scale flow is locally charge neutral (in the sense that $\sigma = 0$, as expected from the local nuclear physics argument) there is no reason to expect this to remain the case  for the filtered flow. In fact, if $\sigma = 0$ we get
\be
\tilde \sigma = - \tilde u_a \langle J^a \rangle \ , 
\ee
effectively the filtered version of \cref{sigsum},
which is not expected to vanish.

This argument only  hints at what a properly developed filtering argument for the electromagnetic problem will entail (we will discuss that problem in more detail elsewhere), but it illustrates the main point we are interested in. The fact that the local charge density vanishes at the equation of state level does not guarantee that this should be true on the evolved scale. In fact, it seems natural to argue that local charge neutrality should not be enforced in a large-scale simulation.

\section{The fluid equations}

Having considered the electromagnetic aspects from different perspectives, it is apparent that the problem is intricate. When it comes to the assumptions associated with magnetohydrodynamics, the low-frequency approximation that leads to the displacement current being ignored (see \cref{dtE}) should be testable on a case by case basis. The issue of charge neutrality is more involved. One would need to consider  how ``local'' the fluid frame has to be for a given flow. This, in turn, imposes a constraint on the scale on which we are able to average over fluid boxes without violating an assumed charge neutrality. Naturally, this has repercussions for the magnetohydrodynamics and---in absence of a convincing argument---one may have to settle for a pragmatic approach. A sensible way forward may be to focus on consistency---making sure that the chosen formulation is internally ``logical''---and then test whether alternative choices make an actual difference.  Before we reflect on the options, we need to consider the fluid dynamics. 

As a model example we consider the  system explored by \citet{2017CQGra..34l5003A}, relevant for the outer core of a  neutron star above the superfluid transition temperature. We assume that  neutrons and  protons are locked to each other as well as any  thermal component (i.e. we ignore heat flow relative to the baryons) but allow the electrons to drift.  This reduces the set-up to a two-fluid problem. We then need conservation laws for baryon number  and total momentum and we also have to account for the charge current. This way we arrive at a system of equations for the baryon number density $\hat n$ and the (Eulerian) fluid velocity $\hat v^i$, the electron number  density $\hat n_\e$ (noting that we also need to be able to work out the charge density $\hat \sigma$) and the charge current $\hat J^i$. 

At this level, we have two coupled fluid degrees of freedom (e.g. associated with $\hat v^a$ and $\hat J^a$), which include  ``plasma'' properties that are often ignored in astrophysical modelling. However, the general system allows the two components to flow with a large relative velocity, which becomes problematic 
when we try to make contact with the microphysics and the equation of state. The reason for this is fairly obvious. Modern equations of state include key many-body interactions but they do so in dynamical equilibrium. The impact of relative flows is rarely considered.

As we have already discussed, the standard approach to making modelling more ``manageable'' involves reducing the problem to a single-fluid one. In magnetohydrodynamics, this reduction follows from \cref{mhdF}. Once the charge current is slaved to the magnetic field, we can ignore the associated dynamics---we only need to track the fluid via the usual Euler equations (adding the Lorentz force). However, suppose  we want to proceed more cautiously (in such a way that we keep better control on the assumptions that led to \cref{mhdF}). What can we then sensibly do to simplify the problem?

The first step is natural (and we have, in fact, already outlined it).
Returning to the evolution problem and the situation from \cref{fibrate}, we can take the fluid frame (represented by $u^a$) and let different fluid components flow according to \cref{uxa}.
Assuming that the linear drift argument holds on this scale (as we have to in order to arrive at a description that does not require the individual $\gamma_\x$ factors, and in turn the individual velocities)  and translating to the point of view of an Eulerian observer, it makes sense to assume that the difference between the two three-velocities $\hat v_\x^a$ and $\hat v^a$ is small, as well. 
Linearizing in the Eulerian velocity difference, we then have
\be
W_\x = (1- \hat v_\x^2)^{-1/2}  \approx W \left[ 1 + W^2 \hat v_a (\hat v_\x^a - \hat v^a ) \right] \ .
\ee
Combining this with
\be
u_\x^a = W_\x \left(N^a +  \hat v_\x^a \right)  \approx \left[  W \left( N^a +  \hat v^a \right) +v_\x^a \right] \ , 
\label{relate}
\ee
we find that 
\be
v_\x^a \approx W\left[\delta^a_b + W^2 \hat v_b ( N^a +\hat v^a)\right] (\hat v_\x^b - \hat v^b ) \ ,
\label{fluidvel} 
 \ee
and it is easy to confirm that the argument is consistent. A small drift in the fluid frame (the left-hand side) corresponds to a small velocity difference according to the Eulerian observer (the right-hand side).

Let us stress the importance of this  intermediate step between the two-fluid model and magnetohydrodynamics, which is (effectively) a one-fluid description. The linear drift allows us to work with a single Lorentz factor, $W$, associated with the bulk flow (the magnitude of which is not restricted by the assumptions). If we do not assume a linear drift we \emph{have to} keep track of the individual velocities (which makes large-scale simulations much more complicated and expensive). 

\subsection{Baryon number conservation}

In general, 
the physical setting we are exploring represents a two-component system with co-moving baryons (and entropy) but allows for a relative charge current. Given this set-up it is natural to associate  the ``fluid velocity'' with the baryons (this is analogous to using the Eckart frame in studies of relativistic heat flow \citep{LivRev} or working with the Favre-weighted frame in the large-eddy context \citep{2021arXiv210701083C}). That is, we have
\be
 \hat v^i = \hat  v_\p^i = \hat v_\n^i \ .
 \label{eckart}
\ee
The foliation approach \citep{2017CQGra..34l5003A} then makes use of Eulerian observers such that (as before)
\be
u^a = W(N^a +  \hat  v^a)  = {W\over \alpha} \left( t^a - \beta^a + \alpha\hat  v^a \right) \ , 
\label{4fol}
\ee
with the same Lorentz factor as before.

Baryon number conservation is ensured by
\be
\partial_t \left( \gamma^{1/2} \hat n\right) +  D_i \left[ \gamma^{1/2} \hat n \left( \alpha  \hat v^i  -\beta^i \right)\right] = 0 \ ,
\label{baryons}
\ee
%\todo{Comment on lapse...}
where the Eulerian number density is related to the co-moving one  by
\be
\hat n = nW \ , 
\ee 
and we have introduced
$  n =    n_\n+  n_\p$. The relation simply encodes the change in number density that arises because of the length contraction due to the relative motion between the fluid and the (Eulerian) observer.

We have expressed \cref{baryons} in the usual flux-conservative form within  the 3+1 approach (as laid out in \citet{2017CQGra..34l5003A}). However, as we have  suggested, when we consider the microphysics it is natural to pay closer attention to the local physics experienced by a family of observers that ride along with the fluid. Then we have (at least) two alternatives. We can choose to describe the physics in the local fluid frame associated with the four velocity $u^a$, or we can try to make the equations look ``similar''  to the more familiar flat space (Newtonian) ones. In this latter approach (see for example \citet{1982MNRAS.198..339T}) one would introduce a global time (associated with $t^a$) and use a spatial tetrad (relative to this time coordinate) to describe the fluid. The fluid  then has four velocity 
\be
u^a = {W \over \alpha} \left( t^a +  V^a \right)  \ ,
\ee
with
\be
V^i = \alpha \hat v^i - \beta^i \ .
\label{fluidframe}
\ee
Making use of this result, we can rewrite \cref{baryons} as
\be
\left( \partial_t + \mathcal L_V \right) \left( \gamma^{1/2} \hat n \right) + \gamma^{1/2} \hat n D_i V^i = 0 \ ,
\ee
or
\be
\partial_t (\gamma^{1/2} \hat n) + D_i(\gamma^{1/2} \hat n V^i)=0 \ , 
\ee
closely resembling the continuity equation from non-relativistic physics. 
In this picture, the linear drift argument involves keeping only linear terms in velocity differences in a frame determined by the global time coordinate. This follows immediately from the relations in the previous section, since
\be
V_\x^a = V^a + \alpha  ( \hat v_\x^a - \hat v^a ) \ . 
\ee
We will use this result later.

\subsection{The fluid frame}

Having discussed the issue of baryon number conservation, we are primed to comment on a question that we have so far avoided. What  exactly do we mean by the ``fluid frame'' when we discuss the local dynamics? The answer involves an element of choice. In fact, in a somewhat underhand manner, we have already introduced one of the options. In the derivation of \cref{baryons} we chose to work in the frame that moves along with the baryons (which were assumed to be locked at the outset). This  gives precise  meaning to the fibration four velocity $u^a$. This choice has the advantage of making the conservation law \cref{baryons} simple and intuitive. To see how this works, consider a general observer $U^a$ within the family of linear drift models. In general, we then have
the baryon flux
\be 
\sum_{\x=\n,\p} n_\x (U^a + v_\x^a)= (n_\n + n_\p) (U^a + v^a) \equiv nu^a \ .
\label{eckframe}
\ee
The argument involves two steps. First  lock the baryons together ($v^a = v_\n^a = v_\p^a$) and then define the co-moving four velocity $u^a$ as the desired frame. The logic is simple but important. Any other choice  introduces a diffusion velocity in the equation for baryon number conservation. 

Now turn to the stress-energy tensor, which takes the form (for the model system we are considering here and leaving out the purely electromagnetic  contribution, see \citet{2021FrASS...8...51A} for more details)
\be 
T^{ab} = \varepsilon U^a U^b + \perp^{ab} p + 2 \sum_\x n_\x \mu_\x U^{(a} v_\x^{b)}
\label{Tab}
\ee
(ignoring quadratic terms in the drift velocities $v_\x^a$), where $\varepsilon$ is the energy density and $p$ is the pressure. We also need the chemical potential $\mu_\x$ for each species. An observer moving along with each individual fluid frame measures the corresponding chemical potential as (introducing tildes to make a distinction at this point)
\be
\tilde \mu_\x = - u^a \mu^\x_a \ .
\ee
If we ignore entrainment (see \citet{LivRev} for the general role of this effect in multifluid systems), then
\be
\mu^\x_a = \mu_\x u^\x_a  
\ee
so we need
\be
\tilde \mu_\x = - \mu_\x ( u^a u^\x_a ) \ .
\ee
Within the linear drift model, it is straightforward to show that $\tilde \mu_\x \approx \mu_\x$. Similarly, if we define the measured number density as
\be
\tilde n_\x = - u_\x n_\x^a 
\ee
then we also have $\tilde n_\x \approx n_\x$. This is important; different fluid observers  agree on both number densities and  chemical potentials, which in turn means that there is no ambiguity associated with issues like chemical equilibrium. 

Returning to  \cref{Tab}, the first two terms on the right-hand side remind us of the perfect fluid result, while the third term represents the energy/momentum flux that arises due to the relative flow. In the example we are considering we lock the neutrons to the protons, but the electrons exhibits a relative flow. That is, we have 
\be 
\sum_\x n_\x \mu_\x U^{(a} v_\x^{b)} = (n_\n \mu_n + n_\p \mu_
\p) U^{(a} v^{b)} + n_\e \mu_e U^{(a} v_\e^{b)}
\label{momflux}
\ee
If we combine this result with the frame choice from \cref{eckframe} then we arrive at an explicit (non-vanishing) expression for the momentum flux. As an alternative, we may use the freedom of choice associated with $U^a$ to ensure that the stress-energy tensor is reduced to the perfect fluid form. In order to do this, we must work in a frame such that 
\be
(n_\n \mu_\n + n_\p \mu_
\p) v^{a} + n_\e \mu_e  v_\e^{a}=0
\label{landframe}
\ee
Implicitly, this prescribes the corresponding four-velocity $u^a$, and corresponds to the standard Landau-Lifshitz frame \citep{LivRev}. 
Finally, once we have defined the frame we can always replace the electron velocity with the charge current, since
\begin{multline}
j^a = \sum_\x q_\x n_\x (U^a + v_\x^a) \\ = e (n_\p - n_\e ) U^a  + e (n_\p v_\p^a - n_\e v_\e^a)\ .
\label{chcurr}
\end{multline}

From this last expression, it is worth noting yet another option. We could introduce the observer frame in such a way that the spatial charge current vanishes. This would involve removing the last term in \cref{chcurr}. However, as this choice would  introduce a drift velocity in the baryon number conservation law as well as a momentum flux in the stress-energy tensor it does not bring any obvious advantages. 

The key conclusion here is that we can introduce the fluid frame in whatever way we find most convenient, with the two options \cref{eckframe} and \cref{landframe} being attractive for different reasons. There is, however, no free lunch. Whichever choice we make, we cannot at the same time arrive at a baryon number conservation law without particle diffusion and a perfect-fluid stress energy tensor. This may be obvious, but it is an important observation as precisely this combination tends to be assumed from the outset in discussions of relativistic magnetohydrodynamics. Hence, the standard results can only be approximately true. In the following we will make the nature of the required approximation precise by adopting the Eckart frame choice and quantifying the ``offending'' momentum terms in the stress-energy tensor. 

\subsection{Energy-momentum conservation}

In order to complete the fluid model, we need the evolution equations for  energy and  momentum. The starting point is the stress-energy tensor from \cref{Tab}, expressed in the 3+1 foliation. After some fairly straightforward algebra, we find that the energy evolves according to
\begin{multline}
 \partial_t  \left( \gamma^{1/2}  \rho \right) +  D_i \left[ \gamma^{1/2} \left( \alpha S^i - \rho \beta^i \right) \right]\\
 =   \gamma^{1/2}  \left( \alpha  S^{ij}K_{ij} - S^i D_i \alpha \right) \ ,
 \label{eneq}
\end{multline}
where, noting the linear drift assumption, we have  
\be
p+\rho = W^2 (p+\varepsilon) + 2 W^4 \sum_\x    \hat v_a (\hat v_\x^a - \hat v^a ) n_\x \mu_\x \ , 
\ee
%\todo{[Need to define pressure and chemical potentials]}
\be
S^i 
=  (p+\rho)  \hat v^i + W^2 \sum_\x  n_\x \mu_\x (\hat v_\x^i - \hat v^ i )  \ , 
\ee
and
\be
S^{ij} = p\gamma^{ij} + S^i  \hat v^j + W^2 \sum_\x n_\x \mu_\x \hat v^i \left( \hat v_\x^j - \hat v^j\right) \ .
\ee 
We also have the momentum equation 
\begin{multline}
\partial_t (\gamma^{1/2} S_i) + D_j \left[ \gamma^{1/2} \left( \alpha S_i^j -S_i \beta^j \right) \right] \\
= \gamma^{1/2} \left( S_j D_i \beta^j - \rho D_i \alpha \right) \ .
\label{moment}
\end{multline}

We need to add the purely electromagnetic contributions---the Lorentz force---to the right-hand side of the fluid equations. This involves
\be
f_\mathrm{L}^b = -  j_a F^{ab} = N^b ( \hat J^a E_a) + ( \hat \sigma E^b + \epsilon^{bad} \hat J_a B_d ) \ ,
\label{flore}
\ee
which means that we need to add, first of all, a term
\be
\alpha \gamma^{1/2} ( \hat J^i E_i) \ , 
\ee
to the right-hand side of \cref{eneq}, representing the electromagnetic contribution to the energy flow and the Joule heating. Secondly, we need a term
\be
\alpha \gamma^{1/2}  ( \hat  \sigma E_i + \epsilon_{ijk} \hat  J^j B^k ) \ , 
\ee
on the right-hand side of \cref{moment}, representing the (spatial) Lorentz force. 

In the case of a charged two-component problem it makes sense to represent the relative flow of the electrons with respect to the baryons by the charge current.  At the linear drift level, we have
\begin{multline}
\hat \sigma = e (\hat n_\p - \hat n_\e ) = e ( W_\p n_\p - W_\e n_\e) \\
= e W \left[ \left(n_\p - n_\e\right) - W^2 n_\e \hat v_a (\hat v_\e^a - \hat v^a ) \right] \ , 
\label{chden}
\end{multline}
and
\begin{multline}
\hat J^a =  e  ( \hat n_\p  \hat  v^a -  \hat n_\e  \hat v_\e^a) 
\approx \hat \sigma \hat v^ a - eW n_\e (\hat v_\e^a-\hat v^a ) \\
\Longrightarrow \quad \hat v_\e^a-\hat v^a \approx { 1\over e W n_\e} \left( \hat \sigma \hat v^ a - \hat J^a \right)  \ .
\label{Jhat}
\end{multline}

At this point, it is natural to recall  the issue of charge neutrality. Inverting \cref{chden} we have
\begin{equation}
    \sigma = e(n_\p - n_\e) = W\left(\hat \sigma - \hat v^a \hat J_a\right) \;, 
\end{equation}
and it is easy to see that if we were to impose the condition that  the system should be charge neutral in the fluid frame,  $n_\p=n_\e$, we must have
\be
\hat v_ a \hat J^a =  \hat \sigma \ ,
\label{neutcon}
\ee
which connects to the earlier discussion of local charge neutrality that led to \cref{neut}. We now see that the condition arises naturally from the multifluid model (at the linear drift level)---a useful consistency check. It is also worth noting that we need to be careful with what is small and what is not (at least not necessarily). In order to remain consistent, we need pay attention to \cref{Jhat}. In general, the linear drift assumption implies that the combination on the right-hand side of \cref{Jhat} must be small, so we should ignore quadratic terms of this form. It does not follow that  $\hat \sigma$ and $\hat J^a$ are individually small. However, if we were to add the assumption of local charge neutrality ($\sigma=0$) then it follows from \cref{chden} that $\hat \sigma$ will be small (of order the linear drift) and hence (via \cref{Jhat}) the spatial current $\hat J^a$ must be small as well.  In essence, whether specific quadratric terms should be ignored in a consistent model  depends on the physics assumptions.

At the linear drift level we  now have
\begin{equation}
    p + \rho \approx (p+ \varepsilon) W^2 + \frac{2 \mu_\e}{e}W^3 \left(\hat\sigma \hat v^2 - \hat v_i \hat J^i\right)
\end{equation}
\be
S^i \approx   (p + \rho)\hat v^i + { \mu_\e W \over e} \left( \hat \sigma \hat v^i - \hat J^i \right) \ ,
\ee
and
\be
S^{ij} \approx  p \gamma^{ij} +  \hat v^i  S^j  + { \mu_\e W \over e}  \hat v^{j} \left( \hat \sigma \hat v^i - \hat J^i \right)  \ .
\ee

Note that there is only one Lorentz factor (associated with the relative velocity between the observer and the fluid frame) in these expressions. 
Moreover, for low velocities, $\hat v\ll c$, we may be able to ignore the $\hat \sigma \hat v^i$ term compared to the charge current.
Also, for weakly relativistic systems one would expect $\mu_\e \approx m_\e \ll m_\b$. The upshot of this is that the terms involving the charge current would be another factor of order 2,000 or so smaller than the other ``fluid'' terms. For this simple reason, these terms tend to be ignored in non-relativistic problems. This  then leads to the usual statement that the relativistic problem follows by adding a perfect fluid stress-energy tensor to the electromagnetic contribution. The validity of this assumption is  less obvious for a neutron star core, where the electron effective mass may be of order 10\% of the baryon (rest) mass ($\mu_\e \approx 100$MeV). In essence, one should consider including the charge current contribution from the outset. At the very least, it would be worthwhile  quantifying its importance by test simulations. 

\subsection{Electron dynamics}

In order to complete the model, we need to keep track of the electron number density (e.g. in order to work out $\mu_\e$) and the charge current. In general, when the fluxes are conserved we have (for each species)
\be
 \partial_t  \left( \gamma^{1/2} \hat n_\x \right) + D_i \left[ \gamma^{1/2} \hat n_\x \left( \alpha \hat v_\x^i - \beta^i \right) \right] = 0 \ .
\label{necon} 
\ee

In the present case  we focus on  the electron number density. The linear drift assumption then leads to 
\begin{multline}
    \hat n_\e = n_\e W_\e \approx n_\e W \left[ 1 + W^2 \hat v_a \left( \hat v_\e^a-\hat v^a \right) \right] 
    \\
    \approx n_\e W\left[1 + \frac{W}{en_\e}\left(\hat\sigma\hat v^2 - \hat v_a \hat J^a\right)\right] \ .
\end{multline}
Moreover, making use of \cref{Jhat} 
we arrive at
\begin{equation}
    \hat n_\e \hat v_\e^i 
    \approx n_\e W \hat v^i + \frac{1}{e} \left( \gamma^i_j + W^2 \hat v^i \hat v_j\right)\left(\hat \sigma \hat v^j - \hat J^j\right) \ .
\end{equation}

As a slight aside, using these results in \cref{necon},  making use of the global time argument and the expression for charge conservation \cref{chargecon}, we find that
the electron fraction $Y_\e = n_\e/n$ satisfies
\be
\left( \partial_t +\mathcal L_V\right) Y_\e =0 \ .
\ee
In essence, the electron fraction is advected by the fluid flow. This  assumption effectively corresponds to situations where the composition of matter remains frozen during the evolution. That is, the relevant nuclear reactions are slow compared to the dynamics of the system (see for example \citet{ham21}). 

The momentum equation for a general component is (correcting a number of typographical errors, basically removing a term involving the extrinsic curvature, tracing back to equations (78)-(80) from \citet{2017CQGra..34l5003A}, and which  propagates through to their equation (129))
\begin{multline}
\left[ \partial_t + (\alpha \hat v_\x^j - \beta^j ) D_j \right]   S^\x_i 
+ S^\x_j D_i \left( \alpha \hat v_\x^j - \beta^j  \right) \\
+  D_i \left[ \alpha \left(  \hat \mu_\x - \hat v_\x^j S^\x_j\right) \right]   
= {\alpha \over \hat n_\x} \mathcal F^\x_i  \ , 
\label{momx}
\end{multline}
where
\be
 \mathcal F^\x_i = e_\x \hat n_\x \left( E_i + \epsilon_{ijk} \hat v_\x^j B^k \right) + \gamma^a_i R^\x_a \ , 
\ee
with the last term representing resistivity \citep{2017CQGra..34l5001A,2017CQGra..34l5002A,2017CQGra..34l5003A}.

Noting that, in the absence of entrainment (which would not normally link electrons and baryons, anyway \citep{LivRev}), we have
\be
S_\x^i = \hat \mu_\x \hat v_\x^i  \ , 
\ee
and recalling that the fluid velocity is $V_\x^i = \alpha \hat v_\x^i - \beta^i$, we see that \cref{momx} can be concisely written:
\be
\left( \partial_t + \mathcal L_{V_\x}\right) S^\x_i + D_i \left( {\alpha \hat \mu_\x \over W_\x^2} \right) = {\alpha \over \hat n_\x} \mathcal F^\x_i  \ . 
\label{momenew}
\ee
 
In the particular case of the electrons we have
\begin{multline}
    \left[ \partial_t + (\alpha \hat v_\e^j - \beta^j ) D_j \right]   S^\e_i  + S^\e_j D_i \left( \alpha \hat v_\e^j - \beta^j  \right) \\ +  D_i \left[ \alpha \left(  \hat \mu_\e - \hat v_\e^j S^\e_j\right) \right]
= {\alpha \over \hat n_\e} \mathcal F^\e_i  
\label{finohm2}
\end{multline}
where (again, correcting an error  in equation (130) from \citet{2017CQGra..34l5003A})
\begin{multline}
    S_\e^i =  \hat \mu_\e  \hat v_\e^i =   \mu_\e W_\e \left[ \hat  v^i  + {1\over e n_\e W} \left(  \hat \sigma \hat v^i -  \hat J^i \right) \right] 
    \\
    \approx \mu_\e W \left[\left(1 + \frac{W}{en_\e}\hat\sigma\right)\hat v^i - \frac{1}{en_\e W}\left(\gamma^i_j + W^2 \hat v^i \hat v_j\right)\hat J^j\right] \ .
\end{multline}

Finally, we need an expression for the resistivity.  From \citet{2017CQGra..34l5001A,2017CQGra..34l5002A,2017CQGra..34l5003A} we have the general result (neglecting reactions) 
\be
\gamma^a_c R^\x_a = \gamma^a_c \sum_{\y\neq\x} \mathcal R^{\x\y} \left(\delta^b_a + v_\x^b u_a \right) (v^\y_b-v^\x_b) \ , 
\ee
where the velocities are with respect to the fluid frame. In the linear drift model, these are related to the Eulerian velocities through \cref{fluidvel}, and 
in the two-component case we are considering we arrive at 
\begin{multline}
    \gamma^a_c R^\e_a = \mathcal R W \left(\delta^a_c +W^2 \hat v^a \hat v_c \right)  \left( \hat v_a - \hat v^\e_a\right) \\
= { \mathcal R \over e n_\e} \left[ \hat J_c - W^2  ( \hat \sigma - \hat v^a \hat J_a) \hat v_c \right] \ .
\end{multline}

\subsection{Ohm's law}\label{subsec:Ohmslaw}

Resistivity is usually implemented at the level of some version of Ohm's law, often viewed as a ``closure condition'' added to the magnetohydrodynamics relation \cref{EMHD}.
In the multi-fluid model, the required relation follows from the electron momentum equation. As a first step, let us assume that we can ignore the electron inertia. Then it follows   from \cref{momenew} that
\begin{multline}
     \mathcal F^\e_i  \approx - e  n_\e W_\e \left( E_i + \epsilon_{ijk} \hat v_\e^j B^k \right) \\ + { \mathcal R \over e n_\e} \left[ \hat J_i - W^2  ( \hat \sigma - \hat v^j \hat J_j) \hat v_i \right] \approx { n_\e W_\e \over \alpha}  D_i \left( {\alpha  \mu_\e \over W_\e} \right)  \ .
\end{multline}

Introducing 
\be
\eta = { \mathcal R \over e^2 n_\e^2 W} \ , 
\ee
and recalling that
\begin{equation}
    W_\e \approx W \left[1 + \frac{W}{en_\e}\left(\hat \sigma \hat v^2 - \hat v_j \hat J^j\right) \right] 
\end{equation}
we have
\begin{multline}
     E_i + \epsilon_{ijk} \hat v^j B^k + \underbrace{\frac{1}{en_\e W}\varepsilon_{ijk}\left(\hat\sigma \hat v^j - \hat J^j\right)B^k}_\mathrm{Hall\ effect} \\ + \underbrace{\frac{1}{\alpha e} D_i \left\{\frac{\alpha\mu_\e}{W} \left[ 1 -\frac{W}{en_\e}\left(\hat \sigma\hat v^2 - \hat v_j \hat J^j\right) \right]\right\}}_\mathrm{chemical} \\= \underbrace{\eta \left[\hat J_i - W^2\left( \hat \sigma - \hat v_j\hat J^j\right)\hat v_i \right]}_\mathrm{resistivity}
     \label{ohms}
\end{multline}
This is the final result, and we have indicated the main features of the model--- the term associated with the Hall drift, that leads to the development of smaller scale features, a ``chemical'' term of the kind that may be related to battery effects (although, as it turns out, not in this case as a pure gradient will not contribute to \eqref{dtB}) and the resistivity. It is rewarding to note that \cref{ohms} is consistent with the text-book result for non-relativistic two-fluid systems, e.g. equation (2.75) in \citet{bellan} (see also \citet{1999stma.book.....M}), once we set $\alpha = W_\e = W \to 1$ (ignoring terms of order $\hat v^2$) and $\hat \sigma \to 0$. 
Depending on the context,  different aspects of \cref{ohms} may or may not be relevant. Hence, it makes sense to consider possible simplifications. First of all, ignoring the term associated with the chemical potential gradient, we have
\begin{multline}\label{ohmhall}
    E_i + \epsilon_{ijk} \hat v^j B^k  + {1\over e n_\e W} \epsilon_{ijk} \left( \hat \sigma \hat v^j - \hat J^j\right)B^k  \\ = \eta \left[ \hat J_i - W^2  \left( \hat \sigma - \hat v^l \hat J_l \right) \hat v_i \right] \ .
\end{multline}
Also leaving out the Hall term, we are left with
\be
E_i + \epsilon_{ijk} \hat v^j B^k   = \eta  \left[ \hat J_i - W^2  \left( \hat \sigma - \hat v^l \hat J_l \right) \hat v_i \right] \ ,
\label{finohms}
\ee
and, finally, if the system is locally charge neutral we have
\be
E_i + \epsilon_{ijk} \hat v^j B^k   = \eta  \hat J_i \ .
\label{hsthree}  
\ee
It is worth noting that, in the absence of resistivity, this relation leads back to the electric field vanishing in the fluid frame (ideal magnetohydrodynamics). This is as expected, but it is nevertheless a useful consistency check.

Through a hierarchy of approximations and simplifications  we have moved from a model that retains the properties of a charged two-component  plasma  to a simple expression encoding Ohm's law. This does not necessarily mean that we are done. We still need to consider how the result can be used in practice.

\subsection{The traditional approach}\label{subsec:tradApproach}

Before we proceed, 
it is instructive to compare the final result \cref{finohms} to the standard argument  \citep{bekor}, which starts from magnetohydrodynamics and arrives at Ohm's law by taking the current to be proportional to the Lorentz force acting on a particle in the fluid frame. Assuming
\be
 \perp_a^b j_b = \kappa F_{ab} u^b \ , 
 \label{ohmp}
\ee
(using $\kappa$ to represent the conductivity to avoid confusion with the charge density $\sigma$) and recalling \cref{fvel}, it readily follows that
\be
 \hat \sigma  + W^2 (\hat v_i\hat J^i - \hat \sigma)     = \kappa W  (\hat v^i E_i) \ , 
 \label{barsigma}
\ee
and
\be
\hat  J_a  - W^2 \hat v_a (\hat \sigma - \hat v_i\hat J^i)
=  \kappa W \left( E_a  + \epsilon_{abc} \hat v^b B^c \right) \ . 
\label{hatj}
\ee
Moreover, we have
\be
\hat v^i \hat J_i - W^2 \hat v^2  (\hat \sigma - \hat v_i\hat J^i)= \kappa W  (\hat v^i E_i) \ , 
\ee
and
\be
E_i  + \epsilon_{ijk} \hat v^j B^k = {1\over \kappa W} \left[ \hat J_i - W^2  (\hat \sigma - \hat v_l\hat J^l) \hat v_i \right] \ . 
\ee
This version of Ohm's law---notably identical to \cref{finohms} once we identify $\eta = 1/\kappa W$ ---has been implemented in recent numerical simulations, see for example \citet{2009MNRAS.394.1727P,2013PhRvD..88d4020D,2020MNRAS.491.5510W}.

\section{A sequence of approximations/assumptions}

We have explored the main aspects of the problem of charged flows in general relativity and the connection with numerical simulations. We considered issues relating to both electromagnetism and the fluid dynamics. The analysis provides everything we need to put together a consistent formulation for magnetohydrodynamics. As such a formulation inevitably involves a number of approximations/assumptions---with different strategies having been adopted in the  literature---it is useful to consider a hierarchy of models of increasing ``simplicity''. 

Let us outline the  main options, 
framing the discussion in the context of neutron star physics---as this is an area where the need for different approaches/approximations is obvious. When we consider the neutron-star problem, it is intuitively clear that electromagnetism in the vacuum region far away from the star must be represented by Maxwell's equations (without local charges or currents). At the same time, the dynamics of the highly conducting degenerate neutron-star  interior can be adequately described within (some version of) magnetohydrodynamics. However, this description becomes problematic close to the star's surface---basically, since the Alfv\'en wave speed diverges as the density vanishes---yet, an immediate transition to vacuum conditions may not be appropriate. Rather, the star's magnetosphere may support a significant effective charge density \citep{1969ApJ...157..869G} and considerable currents. Ignoring the matter inertia in this region one arrives at the force-free assumption \citep{komm,uch,2018PhRvD..98b3010C}, which simplifies the dynamics (albeit bringing its own set of issues to consider). In essence, we inevitably need to consider different---more or less physically distinct---regions. 

\subsection{The top-level model: Dissipative electromagnetism}

At the highest level, it would be natural to consider a model that involves the full dynamics associated with Maxwell's equations combined with a meaningful ``single-fluid'' approximation. Without such an assumption, we would have to consider multi-fluid (plasma) aspects of the problem and these may be associated with both complexity and computational cost, see \citet{2014MNRAS.438..704B,2016MNRAS.458.1939B} for efforts in this direction. As we have seen, the single-fluid reduction involves two steps. First, we need to make the linear-drift assumption. If we do not, then we have to keep track of individual fluid Lorentz factors (which obviously require the individual velocities). Second, we have to neglect the dynamics associated with the charge current (e.g. ignore the electron ``inertia''). This step removes the second fluid degree of freedom, and closes the system through some version of Ohm's law.

As a first step, it is easy to  make contact with ideal magnetohydrodynamics. Starting from, for example,  \cref{barsigma} and \cref{hatj} it is easy to see  that, if $\kappa \to\infty$ (i.e. we have a perfect conductor) we must have $E_i + \epsilon_{ijk} \hat v^j B^k = 0$ so the electric field vanishes in the fluid frame. The problem reduces to the one discussed in \cref{subsec:LocalView} and $\hat J^i$ is slaved to the magnetic field.  The opposite limit is not quite as straightforward. Formally, if $\kappa \to 0$ we get
\begin{equation}
    W^2 (\hat v_i \hat J^i - \hat \sigma \hat v^2) = 0 \;,
\end{equation}
and 
\begin{equation}
    \hat J_i - W^2 \hat v_i (\hat \sigma - \hat v_l \hat J^l) = 0 \;.
\end{equation}
Solving these  two equations we arrive at: 
\begin{align}
    \left(\gamma_{ij} - \frac{\hat v_i \hat v_j}{\hat v^2}\right)\hat J^j = 0\ , \qquad \mbox{and} \qquad
    \hat\sigma = \frac{\hat v_i \hat J^i}{\hat v^2} \;.
\end{align}
If we then decompose $\hat J^i = \hat J_\parallel \hat v^i + \hat J_\perp^i$, the first relation tells us that $\hat J_\perp^i = 0$, while the second leads to $\hat J_\parallel = \hat\sigma$. At the end of the day, in the limit $\kappa \to 0$ we have 
\begin{equation}
    \hat J^i = \hat\sigma \hat v^i \;.
\end{equation}
In principle, this represents a perfect insulator---any charge imbalance (a non-zero $\hat\sigma$) will be carried along with the fluid. If we also assume charge neutrality (cf. \cref{neutcon}), then consistency 
dictates (as we cannot have $\hat v^2 = 1$ for a massive fluid) that $\hat \sigma = 0$ and $\hat J^i = 0$. In this sense, Ohm's law  limits to vacuum electromagnetism. 
    
The next level of complexity is to allow for a finite resistivity/conductivity, without assuming charge neutrality, as in \cref{finohms}. This relation is fairly easy to invert---first dotting \cref{finohms} with $\hat v^i$ to get rid of $\hat v_l\hat J^l$---leading to an  expression for the charge current required to close the system of equations:
\begin{equation}
     \hat J_i = \hat \sigma \hat v_i + \frac{1}{\eta}\Big[\big(\gamma_{ij} - \hat v_i \hat v_j\big)E^j + \epsilon_{ijk}\hat v^j B^k\Big] \;.
\end{equation}
 The result would simplify further if we assumed local charge neutrality, but---as we have discussed at length---it is not clear to what extent this assumption will hold in a real system. 

The inversion required to include the Hall effect is more involved. It can be done, using standard methods from linear algebra (see Appendix~A), but the result is messy (as it mixes dissipative and non-dissipative terms) and perhaps not very instructive. If we leave out the chemical term, i.e. take \cref{ohmhall} as our starting point then we find that the charge current takes the form 
%\begin{multline}
%      \hat{J}^i = {1\over \eta} \Bigg\{ \gamma^{i j} + W^2 \left(\hat{v}^2 \gamma^{i %     + {1\over en_\e \eta W} \bigg[  \epsilon^{i j k} B_k \\+ W^2 (\hat{v}_l B^l) \epsilon^{i j k} \hat{v}_k 
%       + \left( {1\over en_\e \eta W} \right) B^i B^j \bigg] \Bigg\} \\ \left(   E_j  +  \left( 1 + {\hat \sigma \over en_\e W}\right)  \epsilon_{j l m} \hat v^l B^m + \eta W^2 \hat \sigma \hat{v}_j \right) \cr
%      \Bigg\{ 1 + W^2 \hat{v}^2 + \left( {1\over en_\e \eta W}   \right)^2  \left[ B^2 + W^2  \left(\hat{v}_n B^n\right)^2 \right]\Bigg\}^{- 1} \ .
%     \label{inverse}
%\end{multline}
%\todo{or
\begin{multline}
      \hat{J}^i =  \Bigg\{ \gamma^{i j} + W^2 \left(\hat{v}^2 \gamma^{i j} - \hat{v}^i \hat{v}^j\right) 
     + {1\over en_\e \eta W} \bigg[  \epsilon^{i j k} B_k \\+ W^2 (\hat{v}_l B^l) \epsilon^{i j k} \hat{v}_k 
       + \left( {1\over en_\e \eta W} \right) B^i B^j \bigg] \Bigg\} \\ \times \left\{  \hat \sigma \hat{v}_j+ {1\over \eta W^2} \left[  E_j  +  \left( 1 + {\hat \sigma \over en_\e W}\right)  \epsilon_{j l m} \hat v^l B^m \right] \right\} \cr
      \times \Bigg\{ 1 + \left( {1\over en_\e \eta W}   \right)^2  \left[ W^{-2} B^2 +   \left(\hat{v}_n B^n\right)^2 \right]\Bigg\}^{- 1} \ .
     \label{inverse}
\end{multline}
%}
The different effects---the resistivity, local charge density and the Hall effect---are not so easy to isolate from this expression. The physics, which is easy to recognize in \cref{ohms}, has been mixed up. If we want to work at this level, then this is something we have to accept. The expression may be a bit involved, but so be it. 

Formally, we may think of \eqref{inverse}  as  representing a ``tensorial conductivity'' but this should not get confused with the physics of such a mechanism (effectively associated with the fact that it is more difficult for charges to flow across magnetic field lines than along them). In the very simplest case, one would then replace
\cref{hsthree} with something like (see \citet{2009MNRAS.394.1727P} for the analogous expression in the fluid frame)
\begin{equation}
    \begin{split}
    \hat J^i &= \kappa^{ij} \left( E_j + \epsilon_{jkl} \hat v^k B^l \right) \\ 
    &= {1\over \eta} \left( \gamma^{ij} +\kappa_1 \epsilon^{ijk} B_k + \kappa_2 B^i B^j \right)  \left( E_j + \epsilon_{jlm} \hat v^l B^m \right)   
\end{split}
\end{equation}
This kind of expression clearly does not account for the Hall drift (as indicated in \cite{HarutyunyanHall}). Rather, it introduces additional  physics which may be important in its own right. 

\subsection{Resistive magnetohydrodynamics}

So far, we have outlined a fairly general model involving only the assumptions needed to reduce the problem to a single fluid degree of freedom. This is not yet a description of magnetohydrodynamics---at least not in the traditional sense as we kept the electric field in the discussion. In order to bring in the remaining assumptions, it is natural to consider the low-frequency/slow-motion limit of the model\footnote{Note that this does not have to represent the non-relativistic limit. The assumption refers to the timescale associated with the dynamics, not the bulk motion or, indeed, weak gravity.}. The usual argument  then leads to  the displacement current being small in \cref{dtE}, and we arrive at \cref{mhdF}, which provides an algebraic expression for the charge current. 
Crucially, this implies that we should now think of Ohm's law as providing the electric field rather than the charge current. This is advantageous because \cref{ohms} is already written as an expression for the electric field---we do not need an inversion in order to implement the relation (even when we include the Hall drift). We are done.

As a practical  illustration of the result, we can write down the fully relativistic induction equation, including both the charge density and the Hall effect. To do this, we take the pre-Maxwell form of the Ampere law: 
\begin{equation}
    \hat J^i = \frac{1}{\alpha \mu_0 }\epsilon^{ijk}D_j\left(\alpha B_k\right)
\end{equation}
as our starting point. Making use of \cref{ohmhall} this leads to 
\begin{multline}
    E_i = -\left(1 + \frac{\hat \sigma }{e n_e W}\right)\epsilon_{ijk}\hat v^j B^k - \eta W^2 \hat\sigma \hat v_i \\- \frac{1}{en_eW\alpha\mu_0}B^k \left[D_i \left(\alpha B_k\right) - D_k \left(\alpha B_i\right)\right] \\ 
    + \frac{\eta}{\alpha\mu_0} \left(\gamma_{il} - W^2\hat v_i\hat v_l\right)\epsilon^{lmn}D_m\left(\alpha B_n\right) \ .
\end{multline}
Using this in the Faraday \cref{dtB} we arrive at the induction equation
\begin{multline}
\left(\partial_t - \mathcal L_\beta \right) B^i -D_m\left[\left(1+\frac{\hat\sigma}{en_eW}\right)\hat v^i B^m\right] \\ + D_m\left[\left(1+\frac{\hat\sigma}{en_eW}\right)\hat v^mB^i\right] \\
-\epsilon^{ijk}D_j \left(\alpha W^2 \eta \hat\sigma \hat v_k\right) + \epsilon^{ijk} D_j \left[ \frac{1}{en_eW\mu_0}B^l D_l (\alpha B_k)\right] \\+ \epsilon^{ijk}D_j \left[\frac{\eta}{\mu_0}(\gamma_{kl} - W^2 \hat v_k\hat v_l) \epsilon^{lmn}D_m(\alpha B_n)\right] = \alpha K B^i \ .
\label{induct}
\end{multline}
This result is complicated, but it is easy to see how it reduces to something more familiar. First of all, leaving out the Hall term the relation simplifies to
\begin{multline}
\left(\partial_t - \mathcal L_\beta \right) B^i 
-\epsilon^{ijk}D_j \left(\alpha W^2 \eta \hat\sigma \hat v_k\right)\\+ \epsilon^{ijk}D_j \left[\frac{\eta}{\mu_0}(\gamma_{kl} - W^2 \hat v_k\hat v_l) \epsilon^{lmn}D_m(\alpha B_n)\right]\\ -D_m\left(\hat v^i B^m\right) + D_m\left(\hat v^mB^i\right)  = \alpha K B^i \ .
\end{multline}
If we also assume charge neutrality we would have
\begin{multline}
\left(\partial_t - \mathcal L_\beta \right) B^i -D_m\left(\hat v^i B^m\right) + D_m\left(\hat v^mB^i\right)+\\ + D_j\left[\frac{\eta}{\mu_0}D^i(\alpha B^j)\right] - D_j\left[\frac{\eta}{\mu_0}D^j(\alpha B^i)\right]
 = \alpha K B^i
\end{multline}
and it is easy to see that, with $\beta = 0,\,\alpha=1,\,K = 0$ and assuming $\eta $ constant, we end up with the standard textbook (special relativistic) version of the (resistive) induction equation.  It would obviously be interesting to explore to what extent the additional terms in \cref{induct} impact on the large scale magnetic field evolution in a neutron star, but we leave this for future work. 

\subsection{Force-free electrodynamics}

The magnetohydrodynamics approach should be relevant for the dense interior of a magnetized star, while the vacuum Maxwell equations apply at large distances. However, if we want to consider the region immediately outside the star, or indeed the transition through the low-density surface material, then we may need a different prescription. A common assumption is that the magnetosphere is composed of a highly magnetized plasma, which supports a charge current without inducing significant ``fluid'' motion. In essence, there is no significant matter component to balance the Lorentz force and we arrive at what is called force-free electrodynamics. From \cref{flore} we see that we must have
\begin{equation}\label{eq:FFLorentz}
    \hat  \sigma E_i + \epsilon_{ijk} \hat  J^j B^k \approx 0 \; ,
\end{equation}
along with 
\begin{equation}\label{eq:EdotBzero}
    E_i B^i \approx 0 \; ,
\end{equation}
and it also follows that $\hat J^i E_i \approx 0$. The condition of high magnetization requires that 
\begin{equation}\label{eq:MagnetizedPlasma}
    B^2 > E^2\;.
\end{equation}
These three conditions may be taken as the axiomatic ``definition'' of the force-free limit (see e.g. \citet{PaschalidisShapiro} for a discussion of the need for all three conditions.) 

Since the vanishing of the Lorentz force follows from the low-inertia assumption, the force-free model is often described as the low-inertia limit of ideal magnetohydrodynamics \citep{komm,2006MNRAS.368L..30M}. The argument is also motivated by the fact that the last two conditions \cref{eq:EdotBzero,eq:MagnetizedPlasma} hold in ideal magnetohydrodynamics as well. The limit argument may seem intuitive, but it is not quite that straightforward. In ideal magnetohydrodynamics the electric field vanishes in the fluid frame, while  force-free electrodynamics is identified by the electric field vanishing in a frame associated with the charge current. Trivially, the two conditions \cref{EMHD} and \cref{eq:FFLorentz} are identical if we let $\hat v^i \to \hat J^i/\hat \sigma$. That is, the force-free region does not strictly follow from simply taking the  low-inertia limit (at least not in the usual mathematical sense). We do not have an adjustable parameter that takes us from magnetohydrodynamics to the force-free case. The argument relies on a boost of the frame in which the electric field is taken to vanish. With this in mind, it is not surprising that the force-free case does not arise naturally from the equations we have discussed. Our starting point was the two-fluid model and the derivation of, for example, Ohm's law clearly builds on the fluid assumption. As a result, one would not expect to be able to reconcile the derived form of Ohm's law with the conditions in the force-free region (where the plasma can be assumed to be collisionless). Instead, one would have to consider  dissipation associated with collective processes and/or radiation \citep{2004MNRAS.350..427K}, which obviously changes the argument. Having said that, the conditions required for ideal magnetohydrodynamics and force-free electrodynamics are sufficiently similar that it may nevertheless be useful for numerical implementations to link them through a somewhat ad hoc limiting argument \citep{PaschalidisShapiro,PalenzuelaFFErMHD}. 

When it comes to evolving the equations in the force-free region, the typical approach involves solving for both the  electric and magnetic fields. At a glance, we need a ``closure relation'' for the charge current $j^a$ in order to close the system. While this is not exactly true---as we could contract the Maxwell equations with the Faraday tensor, and then use the vanishing of the Lorentz force to get rid of $j^a$ (see \citet{CarrascoFFE16} for discussion)---it might still be useful to write the charge current in terms of $E^i$ and $B^i$ such that the force free constraints are automatically satisfied. This  leads to (see e.g. \citet{2006astro.ph..4364G,Komissarov2011,PaschalidisShapiro})\footnote{This relation is sometimes referred to as ``Ohm's law'' for force-free electrodynamics, but this is clearly misleading as there is no resistivity involved. Instead, the component of the current orthogonal to the magnetic field is easily obtained from \cref{eq:FFLorentz}, while the component along the magnetic field is  obtained by demanding that the evolution preserves the $E^iB_i = 0$ condition.}
\begin{multline}
     \hat J^i = \frac{B^i}{\mu_0 B^2}\Big[ B_j  \epsilon^{jkl} D_k B_l  - E_j \epsilon^{jkl} D_k  E_l \\ - 2 B^j E^k K_{jk}  \Big] 
     +  \frac{\hat\sigma}{B^2}\epsilon^{ijk}E_jB_k \;.
\end{multline}
As  suggested by \citet{PalenzuelaFFErMHD}, we may work with a phenomenological closure for the charge current, such that it limits to ideal magnetohydrodynamics in one case and to force-free  electrodynamics in the other. 

\section{Concluding remarks}

We have explored the physics of general relativistic magnetohydrodynamics, as required for studies of large-scale magnetic field dynamics associated with, for example, neutron star mergers. With this particular application in mind, we formulated the problem using the standard 3+1 foliation approach to spacetime. However, given the need to faithfully represent the physics, we also considered the  spacetime fibration associated with the fluid elements. Our main aim was to discuss commonly made assumptions (which tend to be motivated in the non-relativistic setting and then taken, more or less, for granted in the curved spacetime case) and establish to what extent they are appropriate for different problem settings. 

The discussion brought issues associated with the charge density and charge current into focus, and we highlighted the connection between the microphysics (associated with a given equation of state) and the global dynamics (from the point of view of numerical simulations). This discussion emphasised different effects, that involve going beyond ``standard'' ideal magnetohydrodynamics and which may come into play if a more precise description of the problem is desired. For example, our derivation of Ohm's law takes the two-fluid plasma as its starting point and, hence, includes features beyond the usual scalar resistivity/conductivity (like the Hall effect). This provided a hierarchy of models that should be relevant for future applications.

While we  have (admittedly) not resolved all the involved issues,  the final formulation is consistent, both logically and physically. This  prepares the ground for a new generation of models of various astrophysical scenarios. In particular, our results will allow us to test the validity of different assumptions and simplifications  by direct simulations. This seems like an important step in the right direction.

\section*{Acknowledgments}

Many colleagues have contributed useful discussion during the development of this material. We are particularly grateful to Kiki Dionysopoulou and  Tamanna Jain. We  are also grateful for support from STFC via grant numbers ST/R00045X/1 and ST/V000551/1. 

\appendix 
\section*{Appendix A}\label{app:OhmInversion}

Assuming that each index is raised (lowered) with $\gamma^{i j}$ 
($\gamma_{i j}$)  we can rewrite \cref{ohmhall} in the form (as the chemical term makes this inversion more complicated, we neglect it in the following)
\begin{equation}
 A_{i j} \hat{J}^j = C_i \ , 
\end{equation}
where we have defined
\begin{eqnarray}
    A_{i j} &=& \gamma_{i j} + L \hat{v}_i \hat{v}_j + M \epsilon_{i j k} \hat B^k \ , \\
    C_i &=& N E_i  + P \epsilon_{i j k} \hat v^j B^k + Q \hat{v}_i  \ , \\     
    L &=& W^2 \ , \\
    M &=& \frac{1}{\e n_\e \eta W} \ , \\
    N &=& \frac{1}{\eta} \ , \\
    P &=& \frac{1}{\eta} \left(1 + \frac{\hat{\sigma}}{e n_\e W}\right)\ , \\
    Q &=& W^2 \hat{\sigma} \ . 
\end{eqnarray}
From Linear Algebra we know that the inverse $A^{i j}$ to a $3 \times 3$ matrix $A_{i j}$ is
\begin{equation}
A^{i j} = \frac{\gamma }{2! A} \epsilon^{i k l} \epsilon^{j m n} A_{k m} A_{l n} \ ,
\end{equation}
where $\gamma = \det \gamma_{i j}$ and $A = \det A_{i j}$. The solution is therefore
\begin{equation}
         \hat{J}^i = A^{i j} C_j \ .
\end{equation}
Note that
\begin{eqnarray}
  \epsilon^{i k l} \epsilon_{j m n} &=& 3! \delta^{[i}_j \delta^k_m \delta^{l]}_n \ , \\
  \epsilon^{i k l} \epsilon_{i m n} &=& 2 \delta^{[k}_m \delta^{l]}_n \ , \\  
  \epsilon^{i k l} \epsilon_{i k n} &=& 2 \delta^l_n \ .
\end{eqnarray}
Explicitly, we have
\begin{multline}
     A^{i j} = \frac{\gamma}{2! A} \epsilon^{i k l} \epsilon^{j m n} \left(\gamma_{k m} + L \hat{v}_k \hat{v}_m 
     + M \epsilon_{k m q} B^q\right) \\
      \left(\gamma_{l n} + L \hat{v}_l \hat{v}_n + M \epsilon_{l n r} B^r\right) \\
 = \frac{\gamma}{A} \Big[\gamma^{i j} + L \big(\hat{v}_k \hat{v}^k \gamma^{i j} - \hat{v}^i \hat{v}^j\big) 
     + M \epsilon^{i j k} B_k + \\
     L M \hat{v}_l B^l \epsilon^{i j k} \hat{v}_k 
      + M^2 B^i B^j\Big] 
\end{multline}
and
\begin{equation}
\begin{split}
    A = &\frac{\gamma}{3!} \epsilon^{i k l} \epsilon^{j m n} A_{i j} A_{k m} A_{l n} \\
    = &\frac{\gamma}{3!} \epsilon^{i k l} \epsilon^{j m n} \left(\gamma_{i j} + L \hat{v}_i \hat{v}_j + 
                M \epsilon_{i j p} B^p\right)\\
    &\left(\gamma_{k m} + L \hat{v}_k \hat{v}_m + M \epsilon_{k m q} B^q\right) \left(\gamma_{l n} + L \hat{v}_l \hat{v}_n + M \epsilon_{l n r} B^r\right) \\
    = &\gamma \left[1 + L \hat{v}_n \hat{v}^n + M^2 B_n B^n + L M^2 \left(\hat{v}_n B^n\right)^2\right] \;. 
\end{split}
\end{equation}
The inverse matrix is therefore
\begin{multline}
     A^{i j} = \Big[\gamma^{i j} + L \left(\hat{v}_k \hat{v}^k \gamma^{i j} - \hat{v}^i \hat{v}^j\right) 
     + M \epsilon^{i j k} B_k \\ + L M \hat{v}_l B^l \epsilon^{i j k} \hat{v}_k + M^2 B^i B^j\Big]\\
     \left[1 + L \hat{v}_n \hat{v}^n + M^2 B_n B^n +  L M^2 \left(\hat{v}_n B^n\right)^2\right]^{- 1} \;,
\end{multline}
and, finally, the current density is
\begin{multline}
      \hat{J}^i = \left[\gamma^{i j} + L \left(\hat{v}_k \hat{v}^k \gamma^{i j} - \hat{v}^i \hat{v}^j\right) 
     + M \epsilon^{i j k} B_k + L M \hat{v}_l B^l \epsilon^{i j k} \hat{v}_k 
      \right. \cr
      \left. + M^2 B^i B^j\right] \left(N E_j  + P \epsilon_{j l m} \hat v^l B^m + Q \hat{v}_j \right) \cr
      \left[1 + L \hat{v}_n \hat{v}^n + M^2 B_n B^n + L M^2 \left(\hat{v}_n B^n\right)^2\right]^{- 1} \ .
\end{multline}


\begin{thebibliography}{}
\makeatletter
\relax
\def\mn@urlcharsother{\let\do\@makeother \do\$\do\&\do\#\do\^\do\_\do\%\do\~}
\def\mn@doi{\begingroup\mn@urlcharsother \@ifnextchar [ {\mn@doi@}
  {\mn@doi@[]}}
\def\mn@doi@[#1]#2{\def\@tempa{#1}\ifx\@tempa\@empty \href
  {http://dx.doi.org/#2} {doi:#2}\else \href {http://dx.doi.org/#2} {#1}\fi
  \endgroup}
\def\mn@eprint#1#2{\mn@eprint@#1:#2::\@nil}
\def\mn@eprint@arXiv#1{\href {http://arxiv.org/abs/#1} {{\tt arXiv:#1}}}
\def\mn@eprint@dblp#1{\href {http://dblp.uni-trier.de/rec/bibtex/#1.xml}
  {dblp:#1}}
\def\mn@eprint@#1:#2:#3:#4\@nil{\def\@tempa {#1}\def\@tempb {#2}\def\@tempc
  {#3}\ifx \@tempc \@empty \let \@tempc \@tempb \let \@tempb \@tempa \fi \ifx
  \@tempb \@empty \def\@tempb {arXiv}\fi \@ifundefined
  {mn@eprint@\@tempb}{\@tempb:\@tempc}{\expandafter \expandafter \csname
  mn@eprint@\@tempb\endcsname \expandafter{\@tempc}}}

\bibitem[\protect\citeauthoryear{{Andersson}}{{Andersson}}{2012}]{2012PhRvD..86d3002A}
{Andersson} N.,  2012, \mn@doi [Phys. Rev. D] {10.1103/PhysRevD.86.043002}, {86, 043002}

\bibitem[\protect\citeauthoryear{{Andersson}}{{Andersson}}{2021}]{2021FrASS...8...51A}
{Andersson} N.,  2021, \mn@doi [Frontiers in Astronomy and Space Sciences]
  {10.3389/fspas.2021.659476}, {8, 51}

\bibitem[\protect\citeauthoryear{{Andersson} \& {Comer}}{{Andersson} \&
  {Comer}}{2021}]{LivRev}
{Andersson} N.,  {Comer} G.~L.,  2021, \mn@doi [Living Reviews in Relativity]
  {10.1007/s41114-021-00031-6},  {24, 3}

\bibitem[\protect\citeauthoryear{{Andersson}, {Comer}  \& {Hawke}}{{Andersson}
  et~al.}{2017a}]{2017CQGra..34l5001A}
{Andersson} N.,  {Comer} G.~L.,   {Hawke} I.,  2017a, \mn@doi [Classical and
  Quantum Gravity] {10.1088/1361-6382/aa6b37},  {34, 125001}

\bibitem[\protect\citeauthoryear{{Andersson}, {Dionysopoulou}, {Hawke}  \&
  {Comer}}{{Andersson} et~al.}{2017b}]{2017CQGra..34l5002A}
{Andersson} N.,  {Dionysopoulou} K.,  {Hawke} I.,   {Comer} G.~L.,  2017b,
  \mn@doi [Classical and Quantum Gravity] {10.1088/1361-6382/aa6b3a}, {34, 125002}

\bibitem[\protect\citeauthoryear{{Andersson}, {Hawke}, {Dionysopoulou}  \&
  {Comer}}{{Andersson} et~al.}{2017c}]{2017CQGra..34l5003A}
{Andersson} N.,  {Hawke} I.,  {Dionysopoulou} K.,   {Comer} G.~L.,  2017c,
  \mn@doi [Classical and Quantum Gravity] {10.1088/1361-6382/aa6b39},  {34, 125003}

\bibitem[\protect\citeauthoryear{{Baiotti} \& {Rezzolla}}{{Baiotti} \&
  {Rezzolla}}{2017}]{2017RPPh...80i6901B}
{Baiotti} L.,  {Rezzolla} L.,  2017, \mn@doi [Reports on Progress in Physics]
  {10.1088/1361-6633/aa67bb},  {80, 096901}

\bibitem[\protect\citeauthoryear{{Barkov} \& {Komissarov}}{{Barkov} \&
  {Komissarov}}{2016}]{2016MNRAS.458.1939B}
{Barkov} M.~V.,  {Komissarov} S.~S.,  2016, \mn@doi [MNRAS]
  {10.1093/mnras/stw384},  {458, 1939}

\bibitem[\protect\citeauthoryear{{Barkov}, {Komissarov}, {Korolev}  \&
  {Zankovich}}{{Barkov} et~al.}{2014}]{2014MNRAS.438..704B}
{Barkov} M.,  {Komissarov} S.~S.,  {Korolev} V.,   {Zankovich} A.,  2014,
  \mn@doi [MNRAS] {10.1093/mnras/stt2247},  {438, 704}

\bibitem[\protect\citeauthoryear{{Baumgarte} \& {Shapiro}}{{Baumgarte} \&
  {Shapiro}}{2003}]{baum}
{Baumgarte} T.~W.,  {Shapiro} S.~L.,  2003, \mn@doi [Ap. J.] {10.1086/346103},
   {585, 921}

\bibitem[\protect\citeauthoryear{{Bekenstein} \& {Oron}}{{Bekenstein} \&
  {Oron}}{1978}]{bekor}
{Bekenstein} J.~D.,  {Oron} E.,  1978, \mn@doi [Phys. Rev. D]
  {10.1103/PhysRevD.18.1809}, {18, 1809}

\bibitem[\protect\citeauthoryear{{Bellan}}{{Bellan}}{2006}]{bellan}
{Bellan} P.~M.,  2006, {Fundamentals of Plasma Physics}.
Cambridge University Press, Cambridge

\bibitem[\protect\citeauthoryear{{Bernuzzi}}{{Bernuzzi}}{2020}]{2020GReGr..52..108B}
{Bernuzzi} S.,  2020, \mn@doi [General Relativity and Gravitation]
  {10.1007/s10714-020-02752-5}, {52, 108}

\bibitem[\protect\citeauthoryear{Carrasco \& Reula}{Carrasco \&
  Reula}{2016}]{CarrascoFFE16}
Carrasco F.~L.,  Reula O.~A.,  2016, \mn@doi [Phys. Rev. D] {10.1103/PhysRevD.93.085013} { 93, 085013}

\bibitem[\protect\citeauthoryear{{Carrasco}, {Palenzuela}  \&
  {Reula}}{{Carrasco} et~al.}{2018}]{2018PhRvD..98b3010C}
{Carrasco} F.,  {Palenzuela} C.,   {Reula} O.,  2018, \mn@doi [Phys. Rev. D]
  {10.1103/PhysRevD.98.023010},{ 98, 023010}

\bibitem[\protect\citeauthoryear{{Celora}, {Andersson}, {Hawke}  \&
  {Comer}}{{Celora} et~al.}{2021}]{2021arXiv210701083C}
{Celora} T.,  {Andersson} N.,  {Hawke} I.,   {Comer} G.~L.,  2021, preprint {\em A covariant approach to relativistic large-eddy simulations: The fibration picture}, \href{https://arxiv.org/abs/2107.01083}{gr-qc/2107.01083}

\bibitem[\protect\citeauthoryear{{Cipolletta}, {Kalinani}, {Giangrandi},
  {Giacomazzo}, {Ciolfi}, {Sala}  \& {Giudici}}{{Cipolletta}
  et~al.}{2021}]{2021CQGra..38h5021C}
{Cipolletta} F.,  {Kalinani} J.~V.,  {Giangrandi} E.,  {Giacomazzo} B.,
  {Ciolfi} R.,  {Sala} L.,   {Giudici} B.,  2021, \mn@doi [Classical and
  Quantum Gravity] {10.1088/1361-6382/abebb7},  {38, 085021}

\bibitem[\protect\citeauthoryear{{Dionysopoulou}, {Alic}, {Palenzuela},
  {Rezzolla}  \& {Giacomazzo}}{{Dionysopoulou}
  et~al.}{2013}]{2013PhRvD..88d4020D}
{Dionysopoulou} K.,  {Alic} D.,  {Palenzuela} C.,  {Rezzolla} L.,
  {Giacomazzo} B.,  2013, \mn@doi [Phys. Rev. D] {10.1103/PhysRevD.88.044020},
  {88, 044020}

\bibitem[\protect\citeauthoryear{{Etienne}, {Paschalidis}, {Haas}, {Moesta}  \&
  {Shapiro}}{{Etienne} et~al.}{2020}]{2020ascl.soft04003E}
{Etienne} Z.~B.,  {Paschalidis} V.,  {Haas} R.,  {Moesta} P.,   {Shapiro}
  S.~L.,  2020, {\em IllinoisGRMHD: GRMHD code for dynamical spacetimes}, Astrophysics Source Code Library, 
  (\mn@eprint {ascl} {2004.003})

\bibitem[\protect\citeauthoryear{{Goldreich} \& {Julian}}{{Goldreich} \&
  {Julian}}{1969}]{1969ApJ...157..869G}
{Goldreich} P.,  {Julian} W.~H.,  1969, \mn@doi [\apj] {10.1086/150119},  {157, 869}

\bibitem[\protect\citeauthoryear{{Gruzinov}}{{Gruzinov}}{2006}]{2006astro.ph..4364G}
{Gruzinov} A.,  2006, preprint {\em Force-Free Electrodynamics of Pulsars}, \href{https://arxiv.org/abs/astro-ph/0604364}{astro-ph/0604364}

\bibitem[\protect\citeauthoryear{{Hammond}}{{Hammond} et~al.}{2021}]{ham21}
{Hammond} P.,  {Hawke} I., {Andersson} N.,  2021, preprint {\em Thermal aspects of neutron star mergers}

\bibitem[\protect\citeauthoryear{{Harutyunyan}}{{Harutyunyan} et~al.}{2018}]{HarutyunyanHall}
{Harutyunyan} A.,  {Nathanail} A., {Rezzolla} L., {Sedrakian} A.,  2020, \mn@doi [Eur. Phys. J. A] {10.1140/epja/i2018-12624-1},{ 54, 11, 191}

\bibitem[\protect\citeauthoryear{{Komissarov}}{{Komissarov}}{2002}]{komm}
{Komissarov} S.~S.,  2002, \mn@doi [MNRAS] {10.1046/j.1365-8711.2002.05313.x},
   {336, 759}

\bibitem[\protect\citeauthoryear{{Komissarov}}{{Komissarov}}{2004}]{2004MNRAS.350..427K}
{Komissarov} S.~S.,  2004, \mn@doi [MNRAS] {10.1111/j.1365-2966.2004.07598.x},
   {350, 427}

\bibitem[\protect\citeauthoryear{Komissarov}{Komissarov}{2011}]{Komissarov2011}
Komissarov S.~S.,  2011, \mn@doi[MNRAS]{10.1111/j.1745-3933.2011.01150.x} { 418, 94}

\bibitem[\protect\citeauthoryear{{McKinney}}{{McKinney}}{2006}]{2006MNRAS.368L..30M}
{McKinney} J.~C.,  2006, \mn@doi [MNRAS] {10.1111/j.1745-3933.2006.00150.x},
 {368, L30}

\bibitem[\protect\citeauthoryear{{Mestel}}{{Mestel}}{1999}]{1999stma.book.....M}
{Mestel} L.,  1999, {Stellar magnetism}.
Oxford Unviersity Press, Oxford

\bibitem[\protect\citeauthoryear{Palenzuela}{Palenzuela}{2013}]{PalenzuelaFFErMHD}
Palenzuela C.,  2013, \mn@doi[MNRAS]{10.1093/mnras/stt311} { 431, 1853}

\bibitem[\protect\citeauthoryear{{Palenzuela}, {Lehner}, {Reula}  \&
  {Rezzolla}}{{Palenzuela} et~al.}{2009}]{2009MNRAS.394.1727P}
{Palenzuela} C.,  {Lehner} L.,  {Reula} O.,   {Rezzolla} L.,  2009, \mn@doi
  [MNRAS] {10.1111/j.1365-2966.2009.14454.x},  {394, 1727}

\bibitem[\protect\citeauthoryear{Paschalidis \& Shapiro}{Paschalidis \&
  Shapiro}{2013}]{PaschalidisShapiro}
Paschalidis V.,  Shapiro S.~L.,  2013, \mn@doi[Phys. Rev. D] {10.1103/PhysRevD.88.104031}, { 88, 104031}

\bibitem[\protect\citeauthoryear{{Rezzolla} \& {Zanotti}}{{Rezzolla} \&
  {Zanotti}}{2013}]{2013rehy.book.....R}
{Rezzolla} L.,  {Zanotti} O.,  2013, {Relativistic Hydrodynamics}.
Oxford University Press, Oxford

\bibitem[\protect\citeauthoryear{{Thorne} \& {MacDonald}}{{Thorne} \&
  {MacDonald}}{1982}]{1982MNRAS.198..339T}
{Thorne} K.~S.,  {MacDonald} D.,  1982, \mn@doi [MNRAS]
  {10.1093/mnras/198.2.339}, {198, 339}

\bibitem[\protect\citeauthoryear{{Uchida}}{{Uchida}}{1997}]{uch}
{Uchida} T.,  1997, \mn@doi [Phys. Rev. E] {10.1103/PhysRevE.56.2181}, {56, 2181}

\bibitem[\protect\citeauthoryear{{Wright} \& {Hawke}}{{Wright} \&
  {Hawke}}{2020}]{2020MNRAS.491.5510W}
{Wright} A.~J.,  {Hawke} I.,  2020, \mn@doi [MNRAS] {10.1093/mnras/stz2779},{ 491, 5510}


\makeatother
\end{thebibliography}
\end{document}